\documentclass[12pt,preprint]{aastex}
\received{} 
\accepted{} 
\journalid{}{} 
\articleid{}{} 
\shortauthors{Mitchell, R.C. et~al.}
\shorttitle{Detailed Spectroscopic Analysis of SN 1987A}

\def\ifundefined#1{\expandafter\ifx\csname#1\endcsname\relax}

\newif\ifpdf
\ifx\pdfoutput\undefined
   \pdffalse      
\else
   \pdfoutput=1   
   \pdftrue
\fi

\def\la{\mathrel{\hbox{\rlap{\hbox{\lower4pt\hbox{$\sim$}}}\hbox{$<$}}}}
\def\ga{\mathrel{\hbox{\rlap{\hbox{\lower4pt\hbox{$\sim$}}}\hbox{$>$}}}}

\newcommand{\be}{\begin{eqnarray}}
\newcommand{\ee}{\end{eqnarray}}

\ifundefined{ensuremath}\def\ensuremath#1{\relax\ifmmode{#1}}
\else${#1}$\fi\else\relax\fi
\ifundefined{nuc}\def\nuc#1#2{\relax\ifmmode{}^{#1}{\protect\text{#2}}
\else${}^{#1}$#2\fi}\else\relax\fi

\newcommand{\etal}{et al.}

\newcommand{\kmps}{km~s$^{-1}$}

\newcommand{\msol}{\ensuremath{{\mathrm{M}_\odot}}}

\newcommand{\nni}{\nuc{56}{Ni}}
\newcommand{\xni}{\ensuremath{\mathrm{X}_{\mathrm{Ni}}}}
\def\ang{\hbox{\AA}}

\newcommand{\phx}{\texttt{PHOENIX}}

\newcommand{\phoe}{{\tt PHOENIX}}

\begin{document}

\title{Detailed Spectroscopic Analysis of SN 1987A: The Distance
to the LMC using the SEAM method}

\author{Robert C.~Mitchell\altaffilmark{1}, E.~Baron, David Branch}

\affil{Department of Physics and Astronomy, University of Oklahoma, 440 West
Brooks, Norman, OK 73019-0261, USA}
\email{mitchellrobertc@ambrose.sau.edu, baron@nhn.ou.edu, branch@nhn.ou.edu}

\author{Peter H.~Hauschildt}

\affil{Department of Physics and Astronomy \& Center for Simulational Physics,
University of Georgia, Athens, GA 30602, USA}
\email{yeti@hal.physast.uga.edu}

\author{Peter~E. Nugent}

\affil{Lawrence Berkeley National Laboratory, Berkeley, CA 94720}
\email{penugent@lbl.gov}

\author{Peter Lundqvist}

\affil{SCFAB, Stockholm Observatory, Department of Astronomy, SE-106
91 Stockholm, Sweden} 

\email{peter@astro.su.se}

\author{Sergei Blinnikov}

\affil{Institute of Theoretical and Experimental Physics, 117218,
Moscow, Russia}
\email{blinn@sai.msu.su}

\author{and}

\author{Chun S.~J.~Pun}

\affil{Dept. of Physics, University of Hong Kong, Pokfulam Road, Hong Kong
}
\email{jcspun@hkucc.hku.hk}

\altaffiltext{1}{Present address: Dept of Physics,
St. Ambrose University, 518 W. Locust St., Davenport, IA 52803}

\begin{abstract}
     Supernova 1987A remains the most well-studied
supernova to date. Observations produced excellent broad-band
photometric and spectroscopic coverage over a wide wavelength range at
all epochs. We model the observed spectra from Day 1 to Day 81
using a hydrodynamical model. We show that good agreement can be
obtained at times up to about 60 days, if we allow for extended nickel
mixing. Later than about 60 days the observed Balmer
lines become stronger than our models can reproduce. We show that
this is likely due to a more complicated distribution of gamma-rays
than we allow for in our spherically symmetric calculations. We
present synthetic light curves in $UBVRIJHK$ and a synthetic bolometric
light curve. Using this broad baseline of detailed spectroscopic models
we find a distance modulus of $\mu = 18.5 \pm 0.2$ using the SEAM
method of determining distances to supernovae. We find that the
explosion time agrees with that of the neutrino burst and is
constrained at 68\% confidence to within $\pm 0.9$~days. We argue that the
weak Balmer lines of our detailed model calculations casts doubt on
the accuracy of the purely photometric EPM method. We also suggest
that Type IIP supernovae will be most useful as distance indicators at
early times due to a variety of effects.
\end{abstract}

\section{Introduction\label{intro}}

Observations of SN~1987A produced extensive
broad-band photometry and excellent spectroscopic coverage.
This motivated the development of models to predict its spectral
and photometric evolution. In our analysis here we make use of the
optical data obtained at the Cerro Tololo Inter-American Observatory
(CTIO) \citep{ctio87a88} and the extensive International Ultraviolet
Explorer (IUE) 
data which were obtained, re-reduced and analyzed by \citet{punetal95}.

While SN~1987A led to the growth of observational data, so too did it
lead to an improvement in theoretical models.  \citet{hflh87} modeled
the early phase of SN~1987A, using a pure hydrogen atmosphere and
accounting for Non-Local Thermodynamic Equilibrium (NLTE).
\citet{eastkir89} modeled the first ten days of the SN~1987A
explosion. H~I and He~I were modeled in NLTE, while metal lines were
treated as pure scattering LTE lines.  The pure H/He non-LTE model of
\citet{tak91} resulted in weak Balmer lines, as did a model by
\citet{phhens94}, which used an earlier version of \phoe\ with H~I,
He~I, Mg~II, and Ca~II in NLTE and metal line blanketing in LTE.
These two groups suggested that differences between the theoretical
and actual density distributions might be responsible for the
discrepancy between the synthetic and observed Balmer lines. 

We use the light curve  model of  Blinnikov and
collaborators calculated using the {\tt STELLA} software package
\citep{blinn93j98,blinn87a99a,blinn87a00}.  This model simulates the
light curve for up to six months after the explosion, and includes
allowance for time-dependent, multi-group radiation hydrodynamics,
monochromatic scattering effects, and the effects of spectral lines on
the opacity \citep{blinn87a00}. The light curve model was based
on the stellar evolution models of \citet*{saio88b}, \citet*{saio88a},
and \citet{nhpr88}.  The procedure for synthesizing SN~1987A spectra
from Blinnikov \etal's model and the results are described in
\S~\ref{models}.  In \S~\ref{spectra} we compare our synthetic spectra
to observations. \S~\ref{lc} compares our synthetic light curves with
observations and we derive a distance to SN~1987A. In a final section
we discuss the implications of our results for the use of SNe~II as
distance indicators.

\section{Models\label{models}}

\subsection{\phoe}

\phoe\ is a general radiation transfer code that computes temperature,
opacity, and level populations for each of typically 50--100 radial
zones in a moving stellar envelope.  \phoe\ solves the full NLTE rate
equations and calculates level populations for a multitude of
different atomic species in LTE or NLTE \citep{hbjcam99,short99}.  For
this study, the following species were calculated in NLTE:
H~I(30/435), He~I(19/37), He~II(10/45), O~I(36/66), Ne~I(26/37),
Na~I(53/142), Mg~II(72/340), Si~II(93/436), S~II(84/444),
Ca~II(87/455), Fe~I(494/6903), Fe~II(617/13675), and Fe~III(566/9721),
where e.g. H~I(30/435) indicates that our model atom for H~I
contains 30 levels and 435 permitted transitions treated in full NLTE.

The Blinnikov \etal\ light curve model consists of an ejecta envelope
divided into 300 zones, with compositions defined for each layer
\citep{blinn87a99a,blinn87a00}.  The outer layers of this model were
re-zoned onto the 75--100 zone grid that is used by \phoe, preserving
the compositions in velocity-space.  The radius, density, and
compositions were taken from the Blinnikov \etal\ model and
interpolated onto the new grid. The inner boundary condition was
taken to be diffusive, which is well justified in these
calculations. The total bolometric luminosity provides the outer
boundary condition and we generally take the value from the results of
\citet{blinn87a00}. However since the bolometric luminosity is a
parameter, we study the effects of varying it to improve the fit to
the observed spectra.  After about 18 days from the explosion we
find that the values predicted by the light curve calculations need to
be adjusted in order for the synthetic spectra to match the
observations.  We adjust the bolometric luminosity to find the best
fit to the observed spectra, we do not attempt to use the photometry
as a fitting parameter in any way. All synthetic photometry is an
output of our calculations once we have calculated the synthetic
spectrum. The temperature is determined self-consistently
from the generalized condition of radiative equilibrium.  We use the
initial temperature from Blinnikov \etal\ as our initial guess. The
gamma-ray energy deposition rate, which is the 
energy in gamma-rays deposited in the matter in the envelope per unit
mass per unit time, is calculated for each new zone. We calculate the
gamma-ray deposition function using our self-consistent spherically
symmetric radiative transfer code, assuming a gray opacity for
gamma-rays of $0.06$~cm$^2$~g$^{-1}$ \citep{swsuthhark95,cpk80}.  Our
calculational 
grid was chosen so as to maintain both composition and density
profiles with adequate resolution.

\section{Synthetic Spectra\label{spectra}}

\begin{figure}
\begin{center}
\leavevmode
\includegraphics[width=12cm,angle=90]{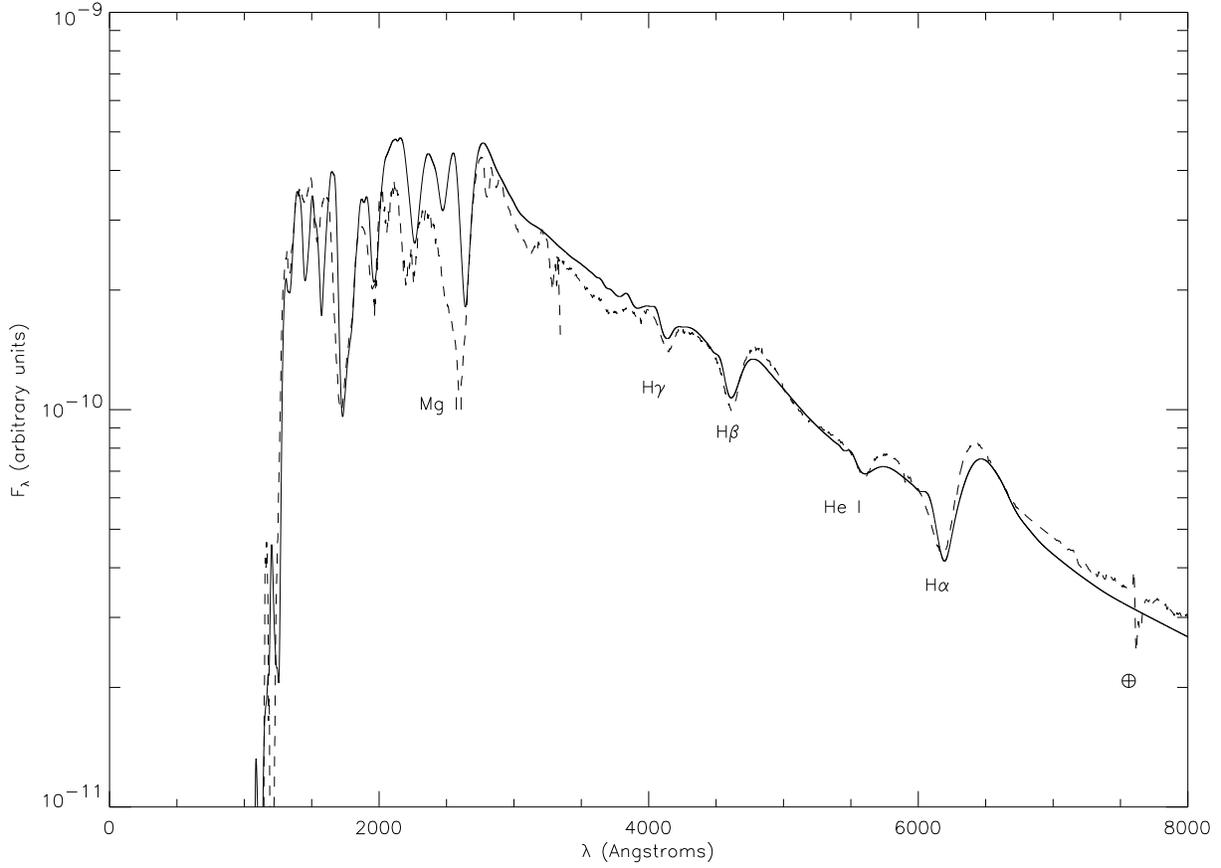}
\end{center}
\caption{\label{fig:1opt} \phx\ model spectrum (solid line) for
day~1.36. Important optical lines include H$\alpha$ through H$\delta$,
and He~I~$\lambda$5876.  In this and following plots, all optical
spectra are taken from the CTIO archive \citep{ctio87a88} and all UV
spectra are from IUE \citep{punetal95}.}
\end{figure}

\begin{figure}
\begin{center}
\leavevmode
\includegraphics[width=14cm,angle=0]{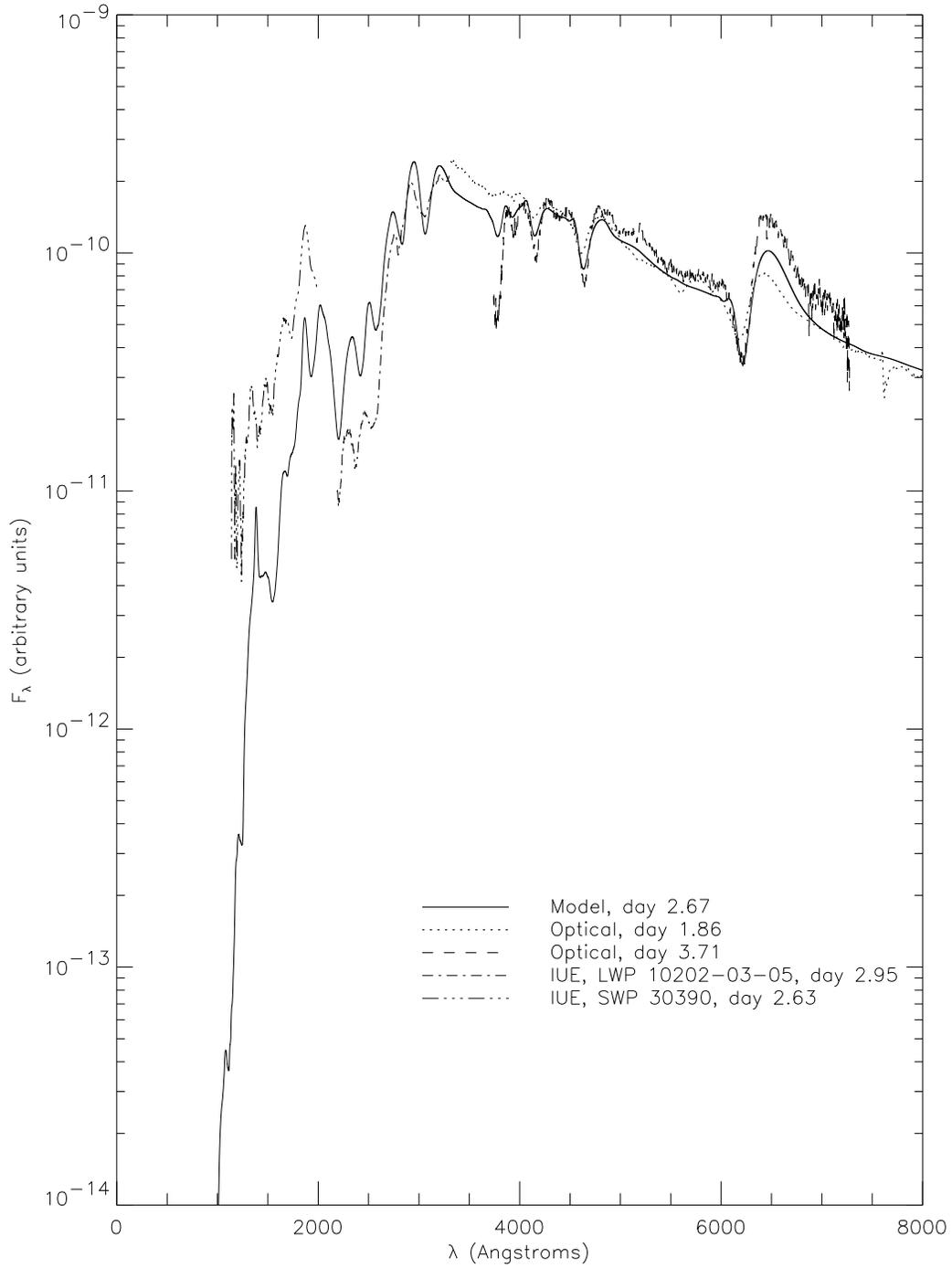}
\end{center}
\caption{\label{fig:2opt} \phx\ model spectrum for day~2.67. In this
and subsequent figures the $1199-1227$~\ang\ wavelength range of the
IUE SWP data are saturated with geocoronal Ly$\alpha$.}
\end{figure}

\begin{figure}
\begin{center}
\leavevmode
\includegraphics[width=14cm,angle=0]{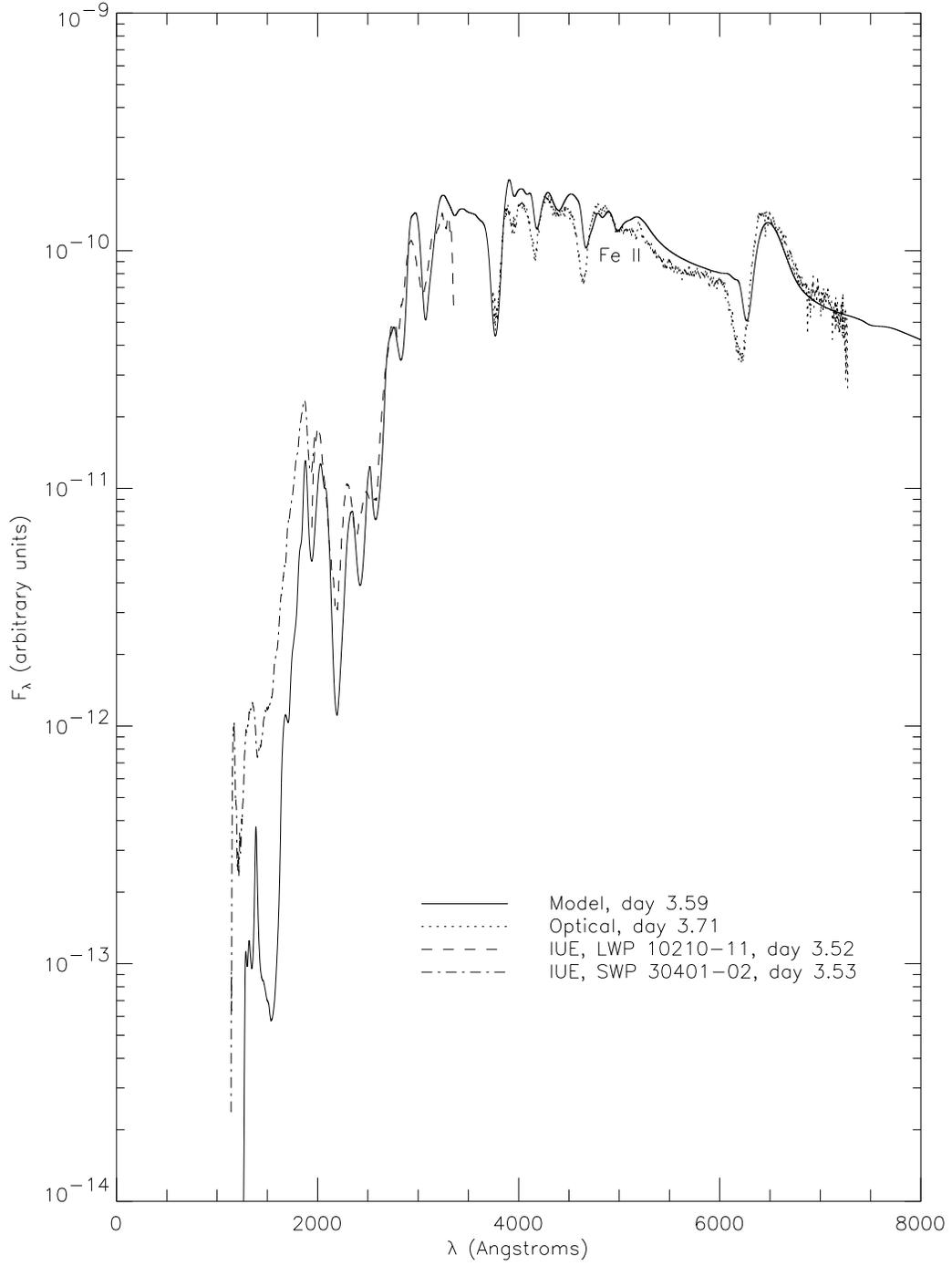}
\end{center}
\caption{\label{fig:3opt} \phx\ model spectrum for day~3.59.}
\end{figure}

Figures~\ref{fig:1opt}--\ref{fig:3opt} show the spectra resulting from
\phoe\ computations on days 1.36, 2.67, and 3.59, respectively, of
SN~1987A.  The fits between the IUE spectra and the synthetic UV
spectra are very good, an indicator of the importance of NLTE effects
in the supernova envelope \citep{snefe296}.  \citet{blinn87a00} noted that the
predicted UV flux 
5--10 days after shock breakout was much greater than the actual flux
when computed in LTE, but this also may have been due to not enough
line blanketing in these initial calculations (E. Sorokina \etal, in
preparation). We included over 1 million lines in all our calculations
and they are dynamically selected to be the most important lines for
the conditions, thus line blanketing is well accounted for in these
calculations. For days 1.36 and 2.67, the
agreement between the synthetic and observed optical spectra is also
reasonably good.  In day 1.36 the blueshifted absorption of the Mg~h+k
line does not extend far 
enough to the blue. The observed feature extends to 43,000~\kmps,
however the hydro model only extends to 33,500~\kmps. Since Mg~h+k
is a resonance line it is optically thick all the way to the surface and
the blue absorption is sensitive to the outermost layers of the
model. Similar results were found by \citet{phhens94} using a
different hydro model and an earlier version of \phx.

\begin{figure}
\begin{center}
\leavevmode
\includegraphics[width=14cm,angle=0]{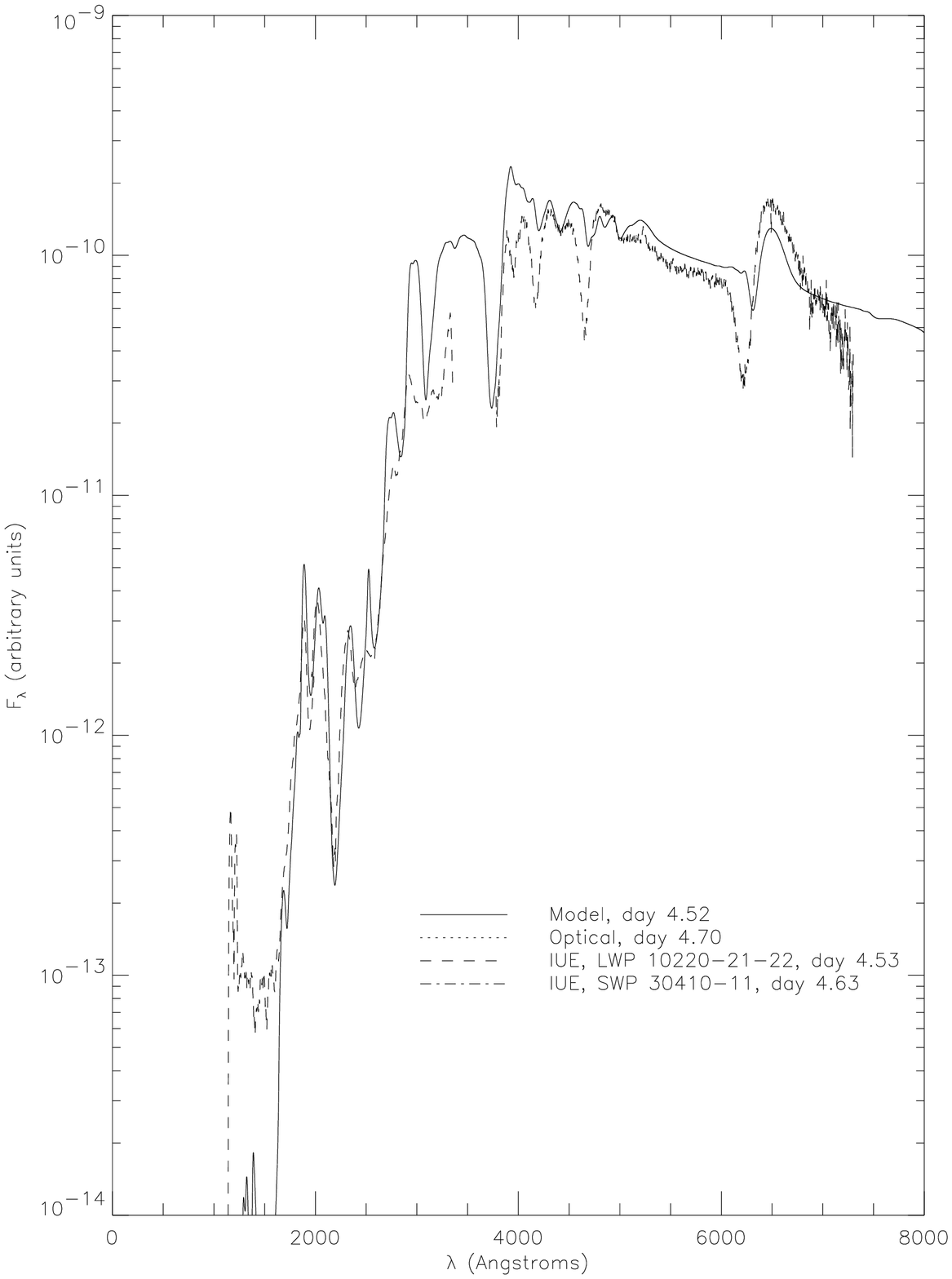}
\end{center}
\caption{\label{fig:4opt} \phx\ model spectrum for day~4.52, without additional
nickel mixing.}
\end{figure}

\begin{figure}
\begin{center}
\leavevmode
\includegraphics[width=14cm,angle=0]{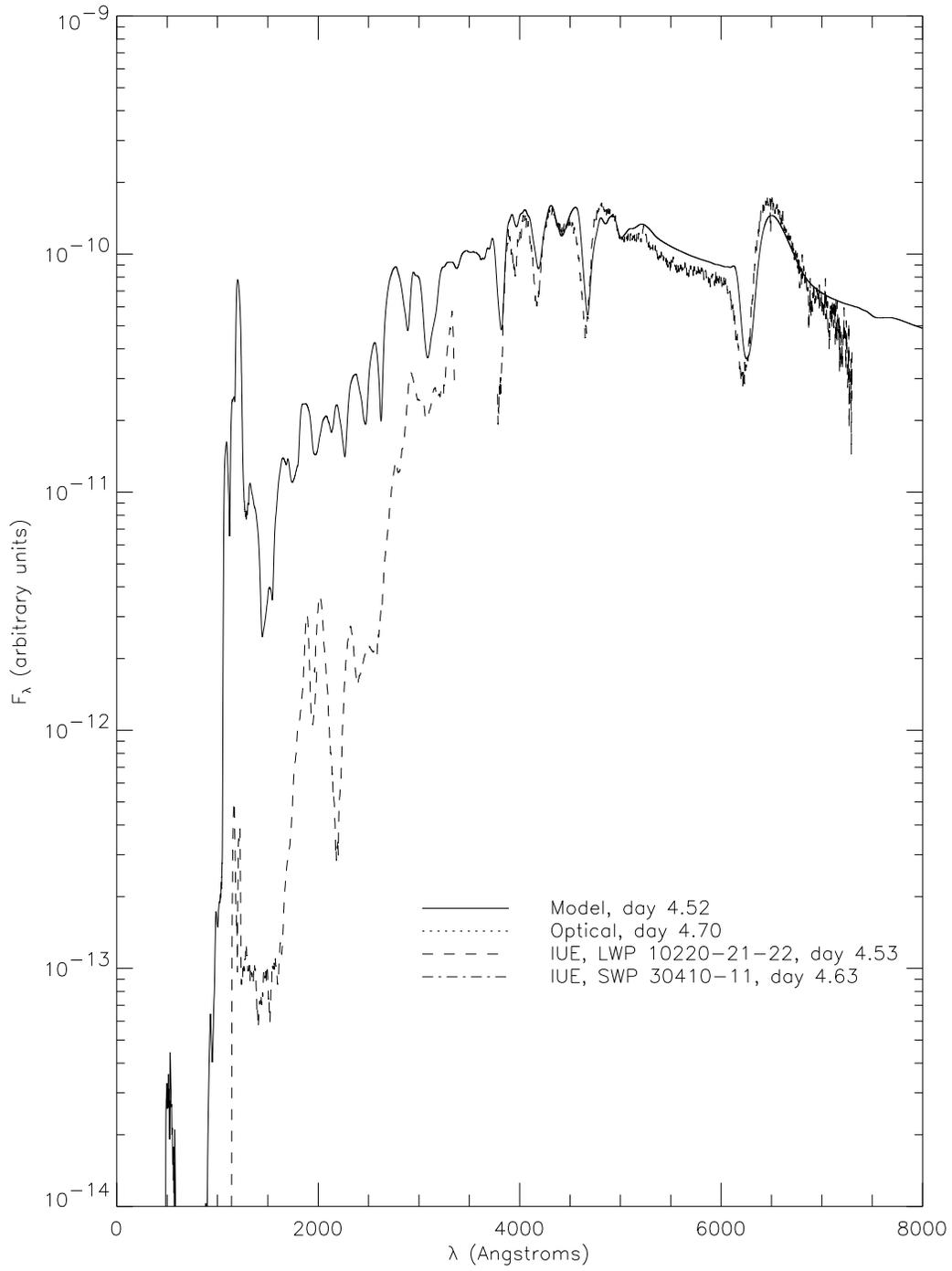}
\end{center}
\caption{\label{fig:7nopt} \phoe\ model spectrum for day~4.52, with gamma-ray
deposition calculated assuming local deposition due to a constant nickel mass
fraction of $1.0 \times 10^{-3}$ everywhere in the envelope.}
\end{figure}

By day 3.59 a discrepancy in the strengths of the Balmer lines begins
to appear in the synthetic spectra, most notably H$\alpha$, even when
NLTE effects are included.  Figure~\ref{fig:4opt} shows the synthetic
optical/near-UV spectrum for day 4.52 compared with the observations
\citep{ctio87a88,punetal95}, and here the Balmer lines are clearly
much weaker in the synthetic spectrum than they are in the observed
spectrum. As we showed in \citet{mitchetal87a01}, we can improve the
fit to the Balmer lines significantly by increasing the gamma-ray
deposition in the outer layers. The result is displayed in
Figure~\ref{fig:7nopt} where we have replaced the self-consistent
gamma-ray deposition function with one that would be obtained from a
uniform nickel mass fraction of $\xni = 1.0 \times 10^{-3}$ in the
envelope.  The fit between the predicted lines and the actual lines is
much better in the optical, but the UV is no longer fit as well. This
shows that the true nickel deposition is more complicated and we have
overestimated the nickel deposition in the very outer layers where the
line-blanketed UV is formed. We have chosen a simple parameterization
to enhance the gamma-ray deposition, i.e., the gamma-ray deposition
follows the density of the model. It is certainly possible (and even
plausible) that there exists a spherically symmetric gamma-ray
deposition function that fits both the optical and the UV, however the
parameter space is large and it is beyond the scope of this work to
find ``the'' correct gamma-ray deposition for each epoch. Nevertheless
our results are a strong indication for enhanced nickel mixing in
SN~1987A. As noted by \citet{mitchetal87a01} the enhanced nickel
mixing improves both the strength of H$\alpha$ and the velocity (as
can be seen by comparing Figs.~\ref{fig:7nopt} and
\ref{fig:4opt}). \citet{UC87A02} find that the velocity and strength
of H$\alpha$ can be well fit by taking into account the effect time
dependence of the hydrogen non-thermal excitation and ionization (the
ionization freeze-out) up to about 30 days, without invoking large
\nni\ mixing.

\begin{figure}
\begin{center}
\leavevmode
\includegraphics[width=14cm,angle=0]{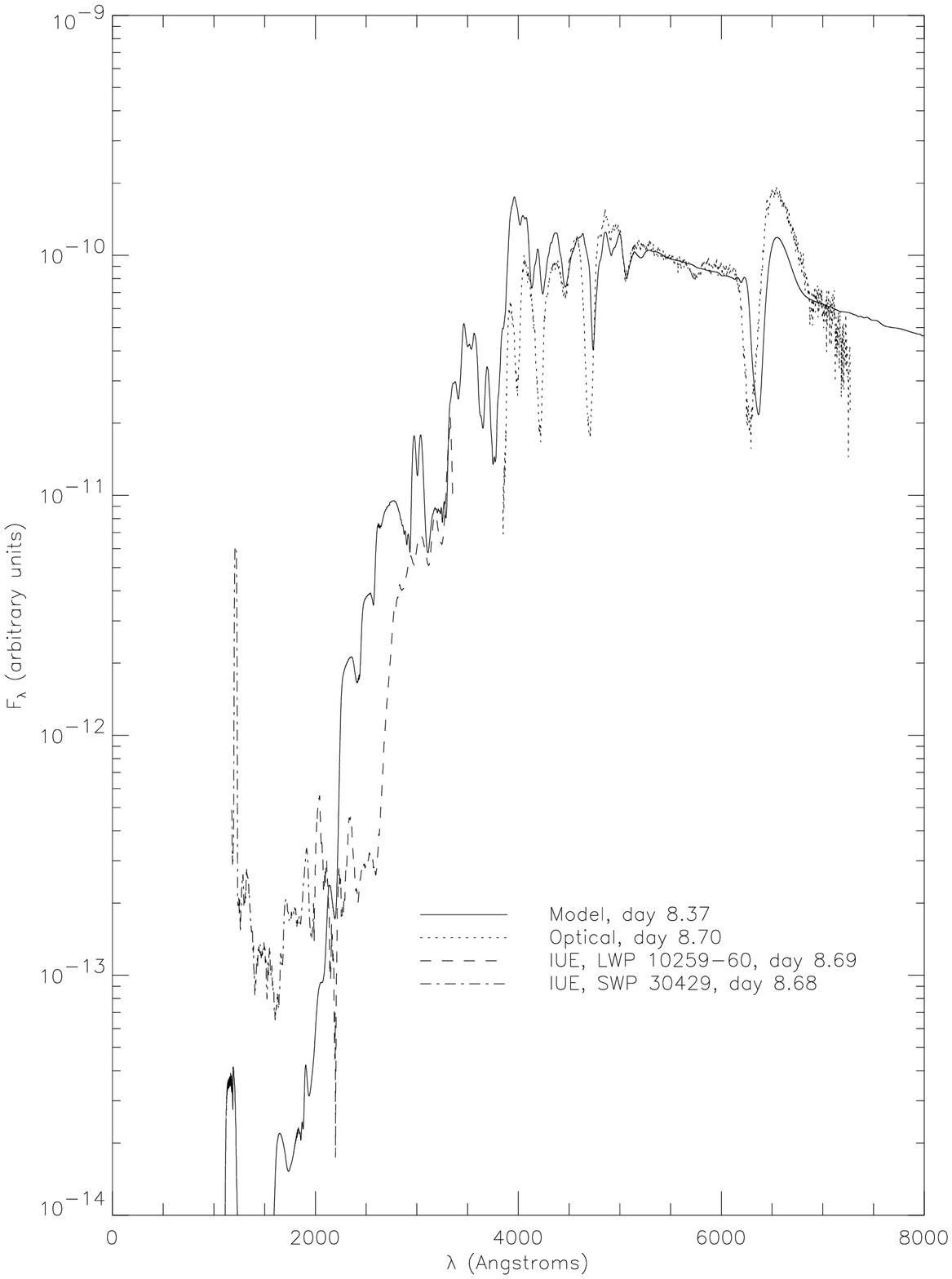}
\end{center}
\caption{\label{fig:8nopt} \phx\ model spectrum for day~8.37, with a
nickel mass 
fraction of $1.0 \times 10^{-4}$ everywhere in the envelope.}
\end{figure}

\begin{figure}
\begin{center}
\leavevmode
\includegraphics[width=14cm,angle=0]{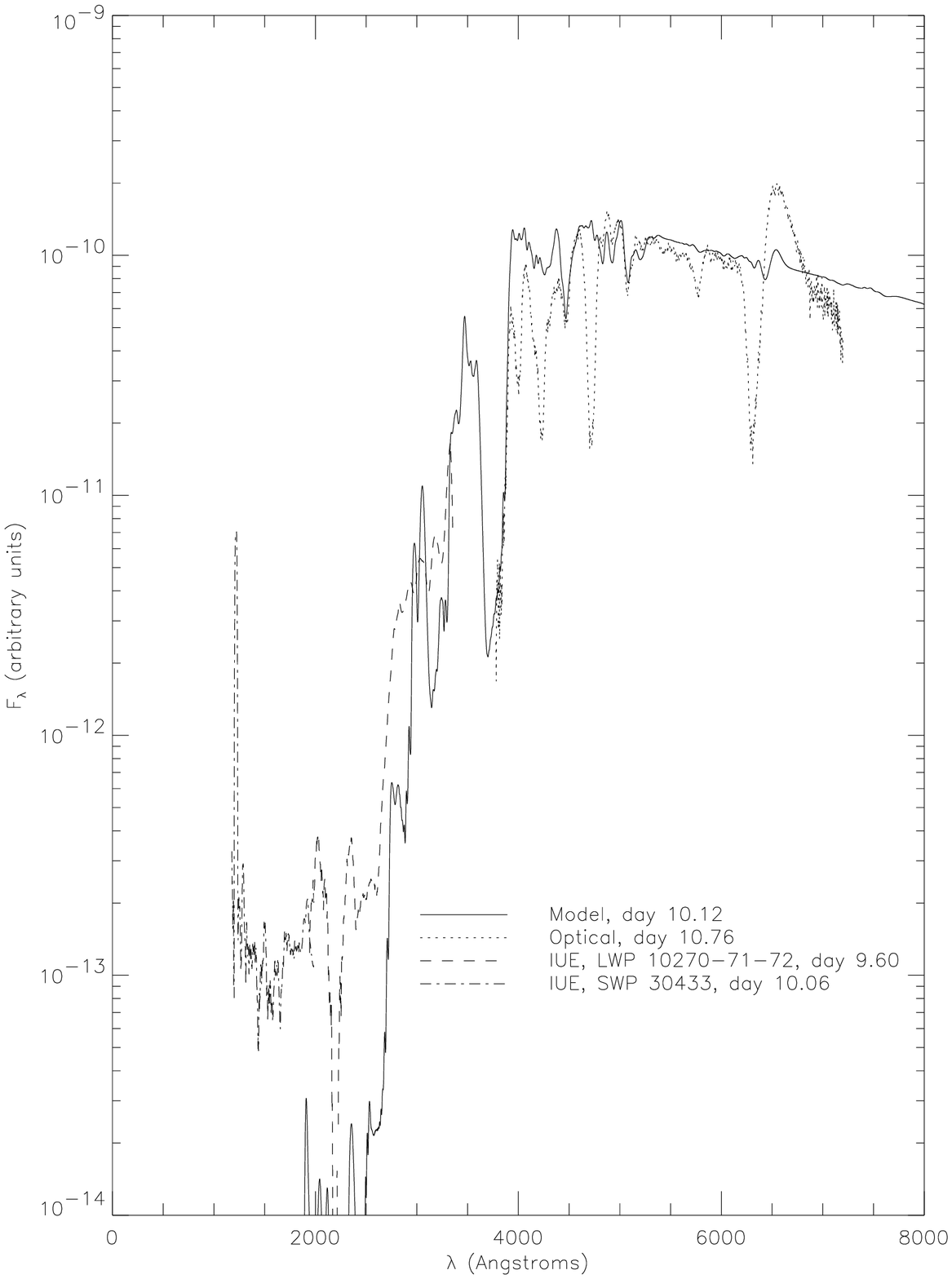}
\end{center}
\caption{\label{fig:10opt} \phx\ model spectrum for day~10.12, without
additional nickel mixing.}
\end{figure}

\begin{figure}
\begin{center}
\leavevmode
\includegraphics[width=14cm,angle=0]{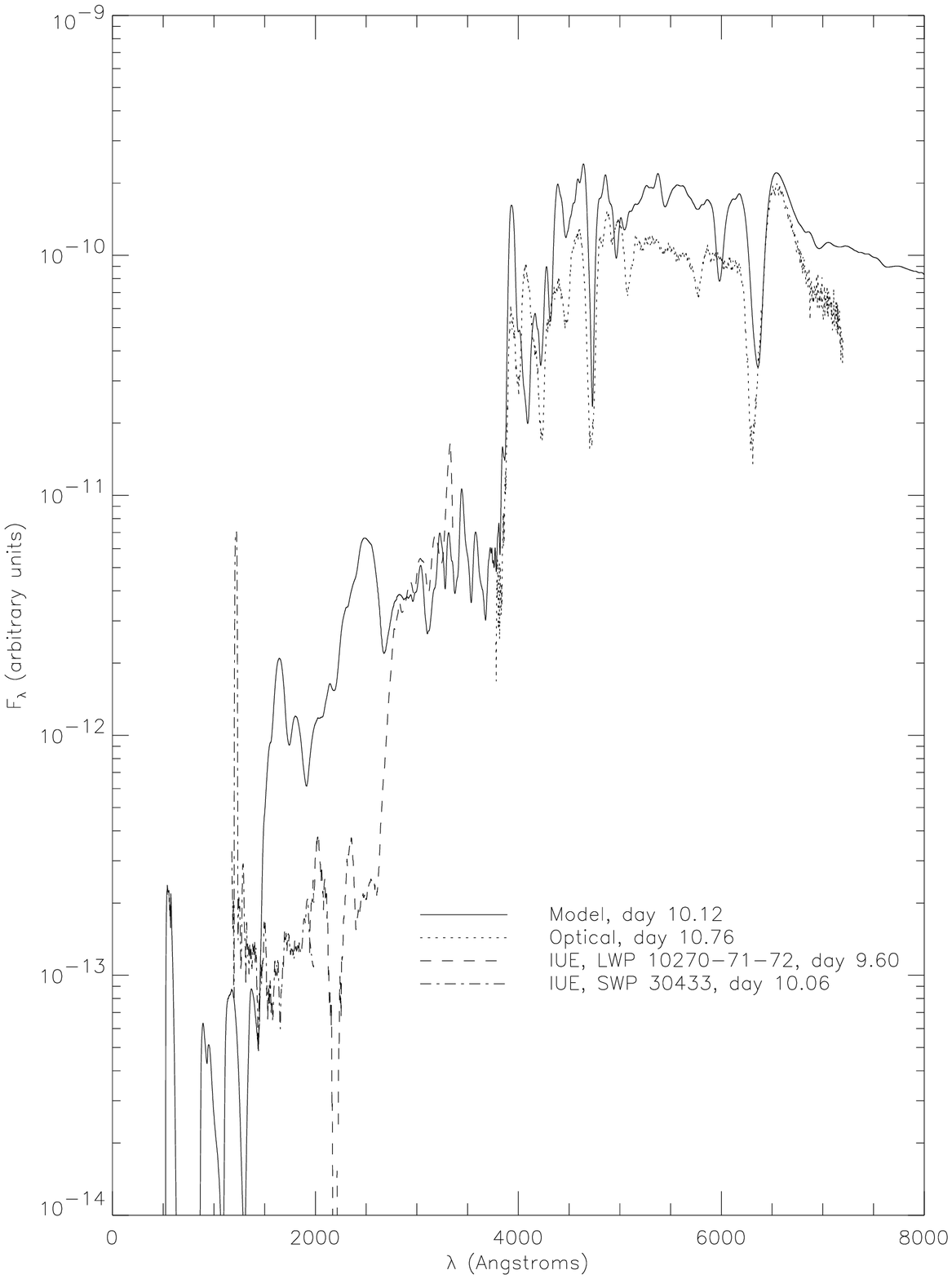}
\end{center}
\caption{\label{fig:10optwni} \phx\ model spectrum for day~10.12, with
additional nickel mixing.}
\end{figure}

\begin{figure}
\begin{center}
\leavevmode
\includegraphics[width=14cm,angle=0]{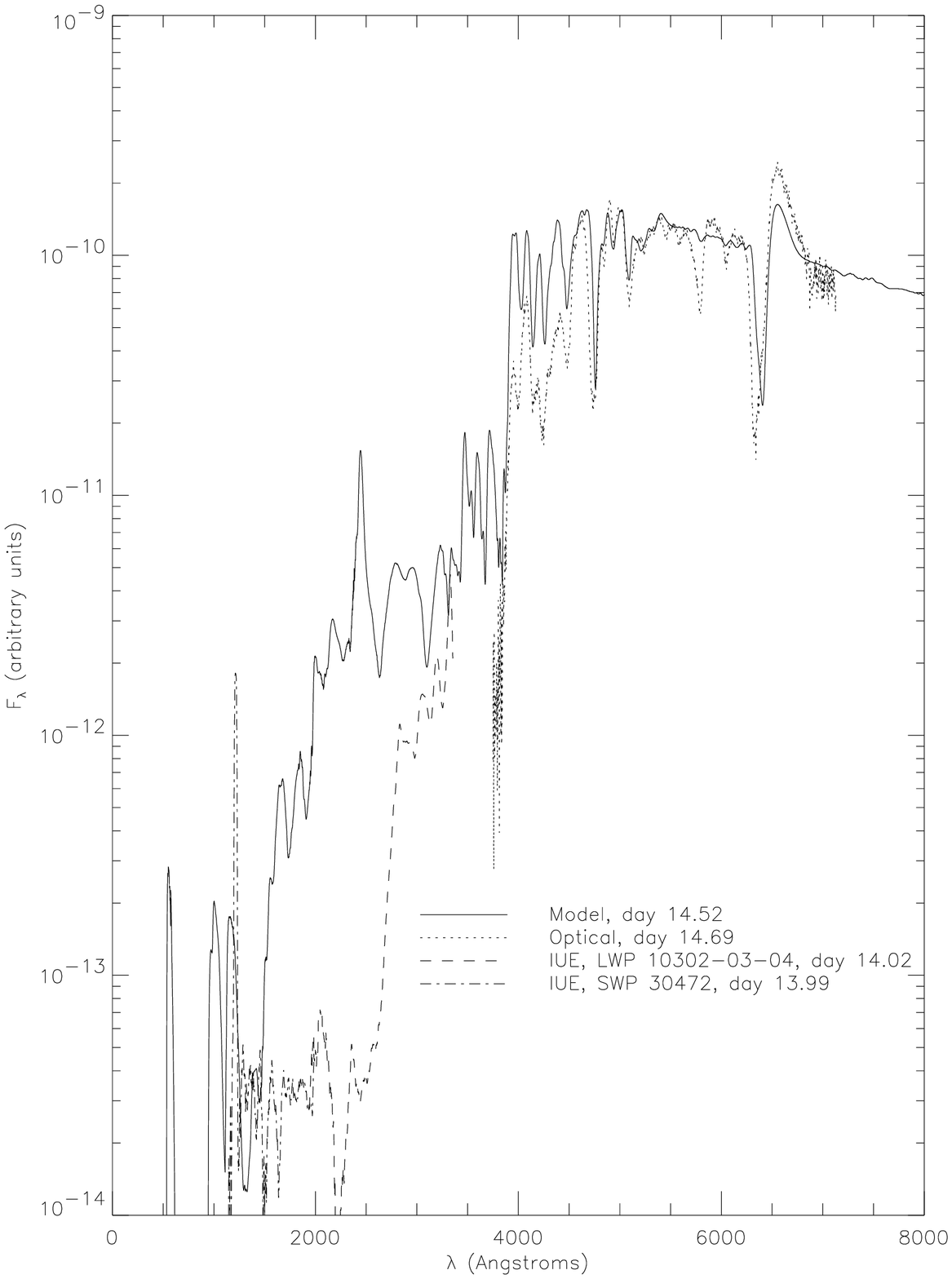}
\end{center}
\caption{\label{fig:14nopt} \phx\ model spectrum for day~14.52, with a nickel
mass fraction of $1.0 \times 10^{-4}$ everywhere in the envelope.}
\end{figure}

Figure~\ref{fig:8nopt} shows the synthetic spectrum compared to the
optical and IUE spectra on day 8. We have assumed uniform nickel
mixing of $\xni = 1.0 \times 10^{-4}$. Comparing this to the day 4
spectrum, we can see that there is not quite enough nickel mixing to
produce the Balmer lines, but the UV won't allow for more mixing in
the simple parameterization we employ. Figure~\ref{fig:10opt} shows the
day 10 spectrum with no additional nickel mixing and again the Balmer
lines are far too weak, and only the overall color in the optical is
reproduced. Figure~\ref{fig:10optwni} shows that nickel mixing
improves the blue and the strength of the Balmer lines, however the
overall fit is not better. Figure~\ref{fig:14nopt} shows the day 14
spectrum, with nickel mixing assumed. The overall fit is not bad in the
optical, but again with the too simple deposition function the UV fit
is poor. The synthetic spectrum shows a weak feature that has been
identified as \ion{Ba}{2} \citep[see][]{williams87A87,jb90}, but since
s-process enhancements have not been included in the stellar evolution
model we use we don't expect to reproduce the observed strength of
this or other s-process elements.
The \ion{Na}{1} 
feature is far too weak in the synthetic
spectrum. 

\begin{figure}
\begin{center}
\leavevmode
\includegraphics[width=14cm,angle=0]{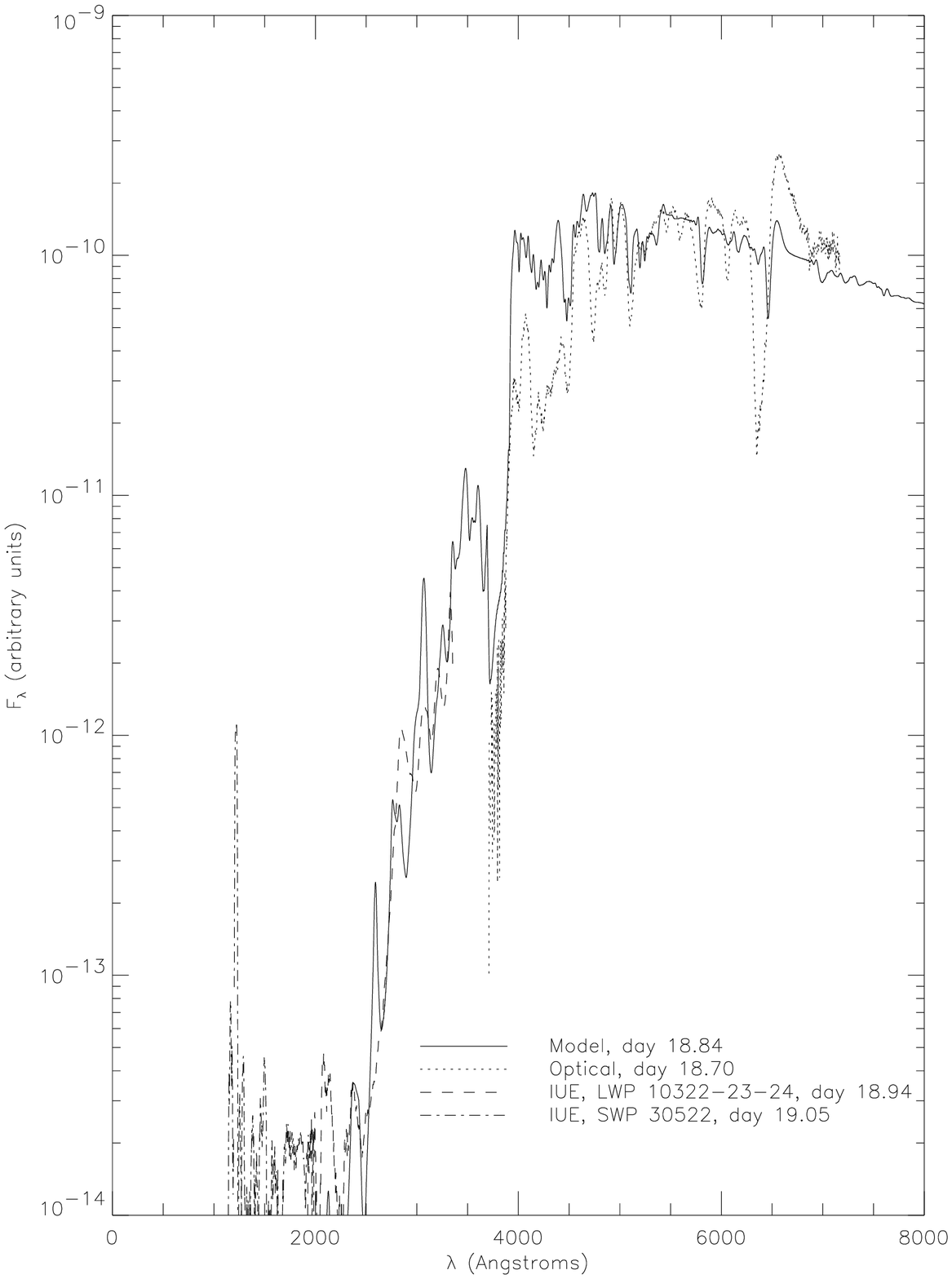}
\end{center}
\caption{\label{fig:18lopt} \phx\ model spectrum for day~18.70, with a nickel
mass fraction of $1.0 \times 10^{-4}$ everywhere in the envelope,
and an increase in bolometric luminosity of 25.6\% from the \citet{blinn87a00}
original value.}
\end{figure}

\begin{figure}
\begin{center}
\leavevmode
\includegraphics[width=14cm,angle=0]{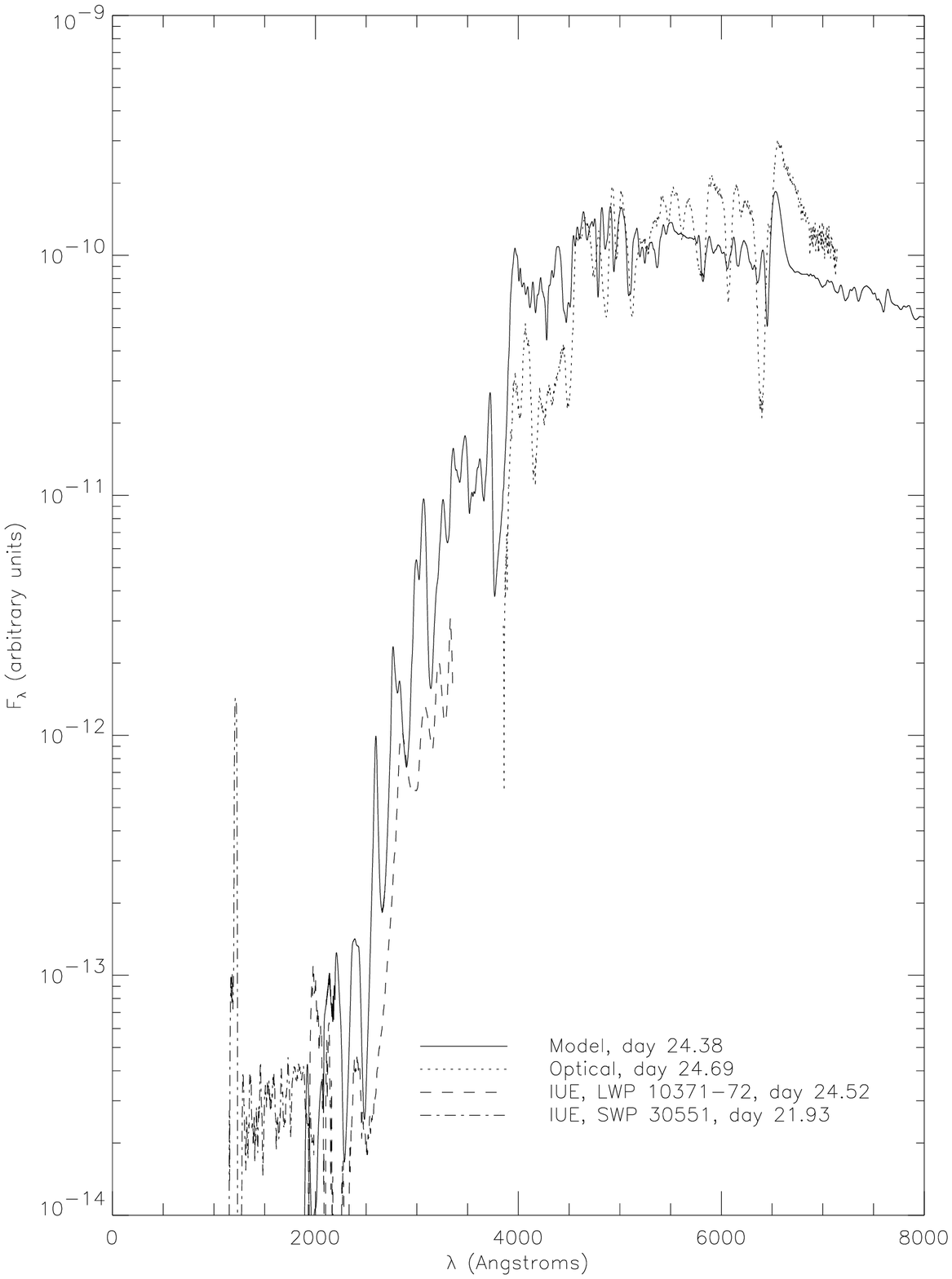}
\end{center}
\caption{\label{fig:24lopt} \phx\ model spectrum for day~24.38, with a nickel
mass fraction of $1.0 \times 10^{-4}$ everywhere in the envelope,
and an increase in bolometric luminosity of 30.9\% from the \citet{blinn87a00}
original value.}
\end{figure}

\begin{figure}
\begin{center}
\leavevmode
\includegraphics[width=14cm,angle=0]{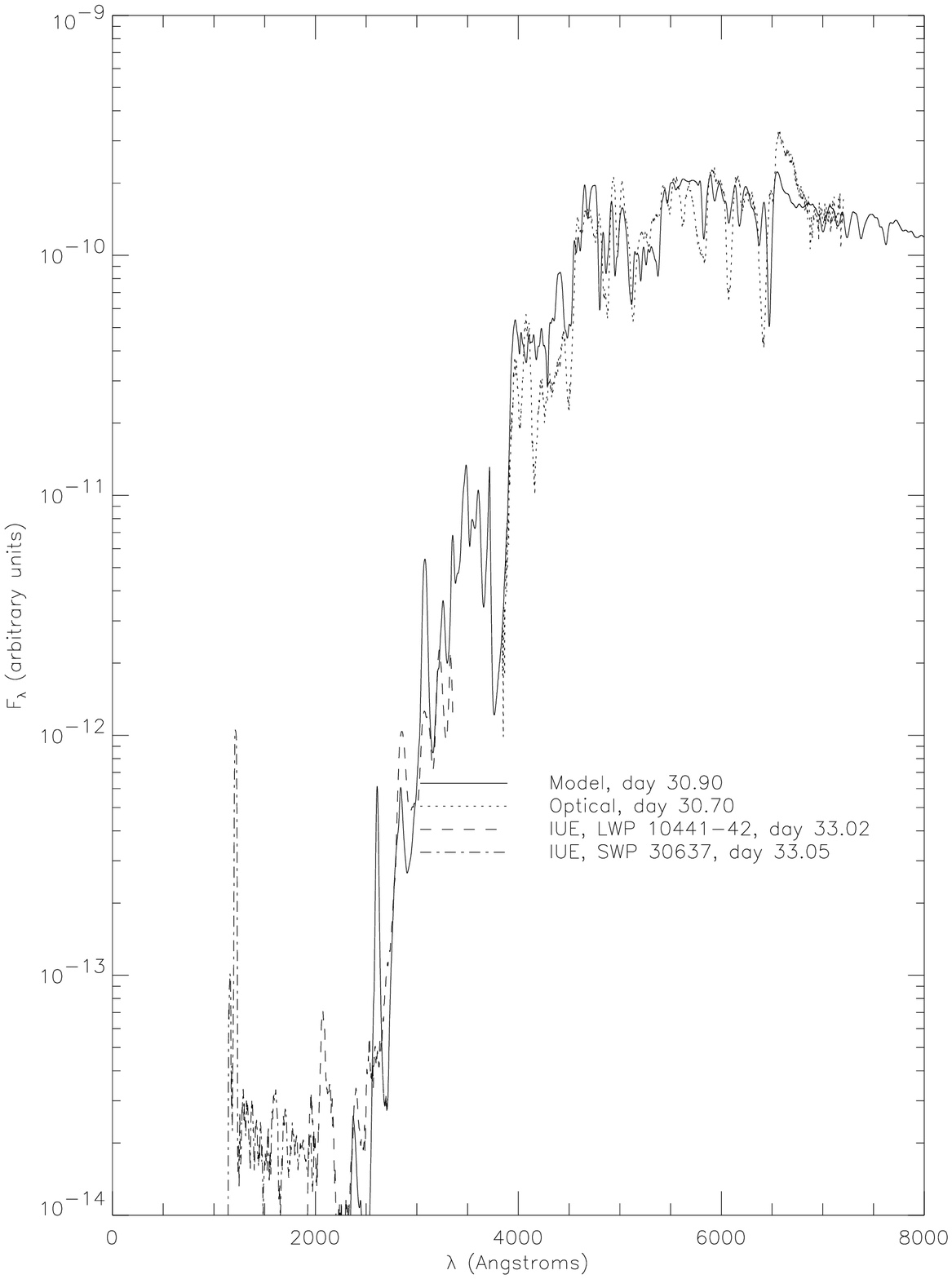}
\end{center}
\caption{\label{fig:30lopt} \phx\ model spectrum for day~30.90, with
an increase in bolometric luminosity of 36.0\% from the \citet{blinn87a00}
original value.  No additional nickel mixing.}
\end{figure}

\begin{figure}
\begin{center}
\leavevmode
\includegraphics[width=14cm,angle=0]{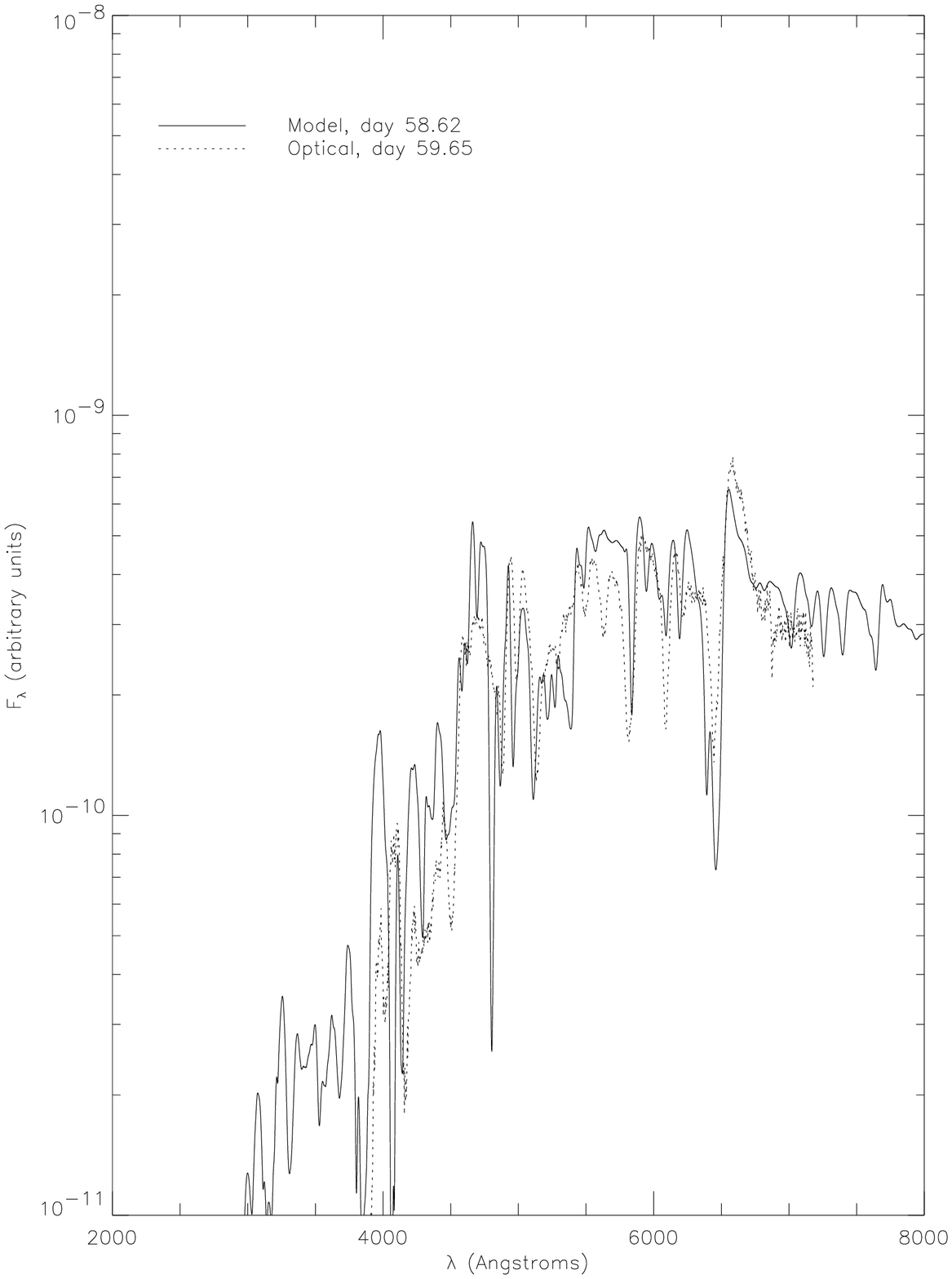}
\end{center}
\caption{\label{fig:58lopt} \phx\ model spectrum for day~58.62, with
an increase in bolometric luminosity of 28.0\% from the \citet{blinn87a00}
original value.  No additional nickel mixing.}
\end{figure}

Figure~\ref{fig:18lopt} shows the synthetic and observed
spectra of day 19, where 
we have both assumed uniform mixing with $\xni = 1.0 \times 10^{-4}$,
and we have increased the bolometric luminosity (a boundary condition
for our modeling) of the model by 26\%
over that obtained from the light curve calculations. We find that if
we simply use the bolometric luminosity of \citet{blinn87a00} we can
not reproduce the observed spectra, but the variations we employ are well
within the errors of \citet{blinn87a00} (see their figure 15). We have
performed a series of calculations where the bolometric luminosity is
varied and only present the best fits here. The best fits are obtained
using the spectra only, no attempt was made to use photometric colors
or any photometric data. The synthetic photometry is an \emph{output}
of our calculations. Except for a
significant over-prediction of the flux just redward of Ca H+K, the
fit is not bad. The Balmer lines are mostly too weak, but the
Na D feature is significantly stronger than just 4 days
earlier, even though we have increased the target luminosity of the
model over that obtained from the light-curve
calculation. Figure~\ref{fig:24lopt} shows the spectra at day 24,
where again we assume constant nickel mixing and have increased the
bolometric luminosity by 31\%. Figure~\ref{fig:30lopt} shows the
spectra at day 31 and the fit from the near UV to the optical is quite
good. The flux is too low compared to that in the \emph{IUE} SWP band. Even
without additional nickel mixing the H$\alpha$ line is reasonably
reproduced.  Figure~\ref{fig:58lopt} shows the optical spectrum at day
58 and again the H$\alpha$ line is well reproduced. 
The structure of the
Balmer lines is in fact dependent not only on the amount of nickel
mixing, but also on the details of the density structure as the
pseudo-photosphere recedes in mass. We have performed a series of
test calculations to verify this (P.~Nugent \etal, in preparation). 
In the observations H$\beta$ is blended with an \ion{Fe}{2} line just
to the red, in the models the lines are unblended and H$\beta$ appears
too strong. This may be due to the nickel mixing in the supernova,
which leads to stronger blending of the iron than is present in the model.

\begin{figure}
\begin{center}
\leavevmode
\includegraphics[width=14cm,angle=0]{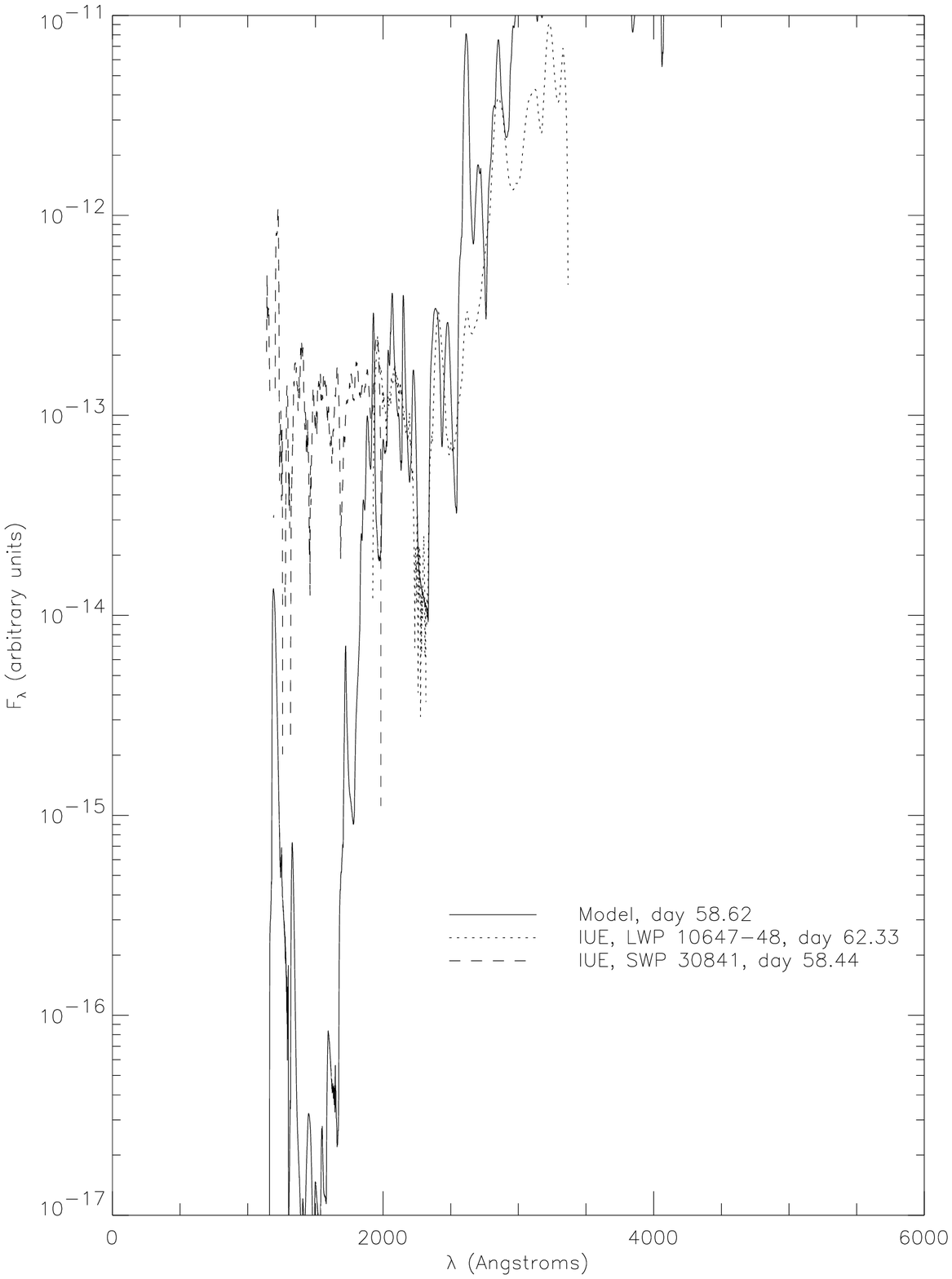}
\end{center}
\caption{\label{fig:58luv} The UV portion of \phx\ model spectrum for
day~58.62, with an increase in bolometric luminosity of 28.0\% from
the \citet{blinn87a00} original value.  No additional nickel mixing.}
\end{figure}

\begin{figure}
\begin{center}
\leavevmode
\includegraphics[width=14cm,angle=0]{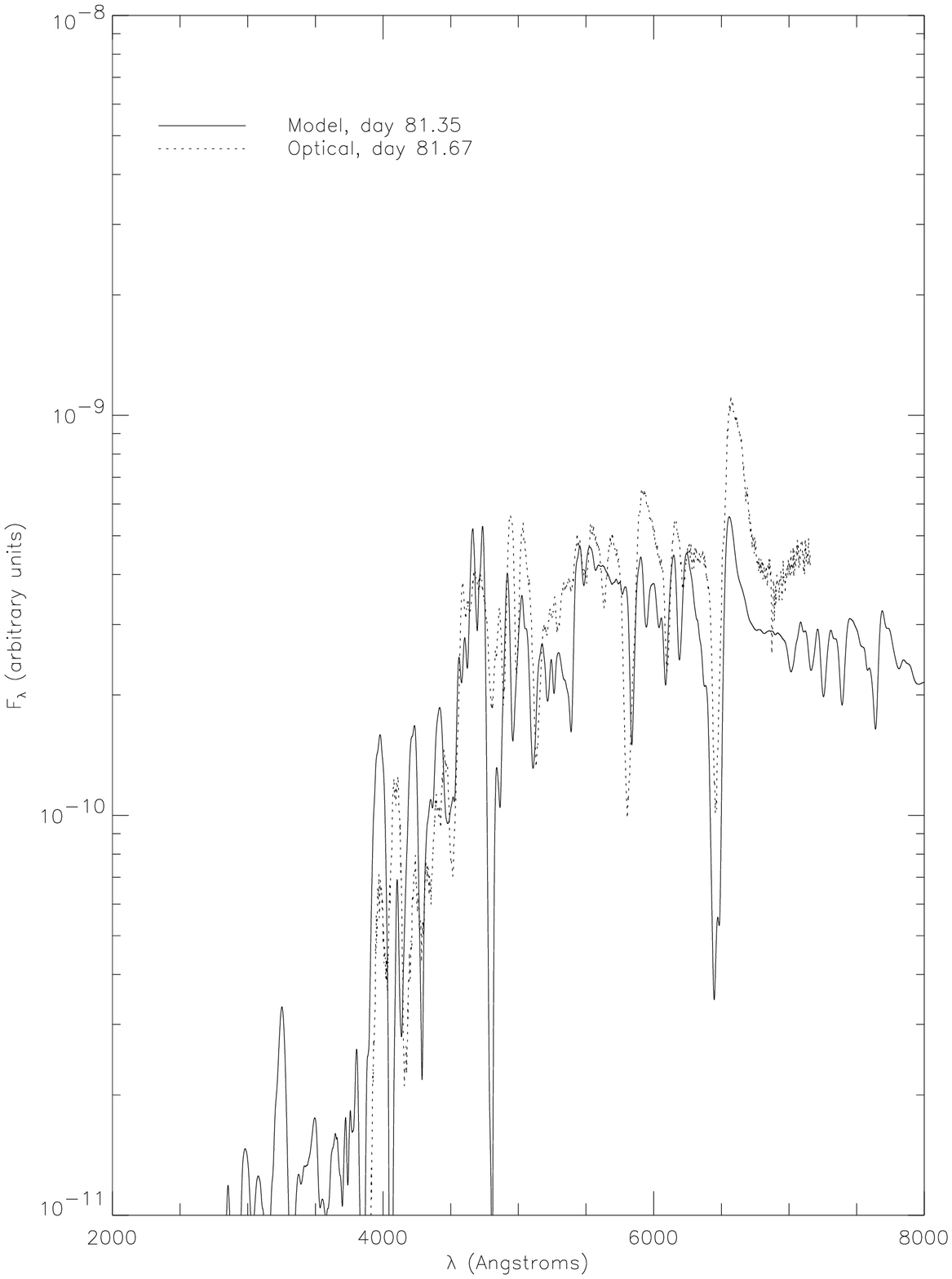}
\end{center}
\caption{\label{fig:81lopt} \phx\ model spectrum for day~81.35, with
an increase in bolometric luminosity of 17.5\% from the \citet{blinn87a00}
original value.  No additional nickel mixing.}
\end{figure}

Figure~\ref{fig:58luv} shows that the
LWP portion of the \emph{IUE} spectrum is well-fit, however the SWP is
not. Figure~\ref{fig:81lopt} shows the day 81 spectra with an increase
in the bolometric luminosity of 17.5\% over the light-curve value.

We have found that the calculated bolometric luminosity in the
lightcurve models is too low at some epochs. We also find that the
assumed mixing of nickel is 
also too small, particularly at early times. Our simple enhanced
gamma-ray deposition (that the gamma-ray deposition follows the
density) does not contain enough parameters to fit both the blue and
red portions of the observed spectra. We note that the synthetic
spectra are sensitive to the \emph{rate} of gamma-ray deposition as a
function of position in the ejecta. At early times almost the entire
envelope is optically thick to gamma-rays and therefore we can
estimate the total mass of nickel in the outer layers
\citep{mitchetal87a01}. At later times 
gamma-rays are transported from the inner layers into the outer shell,
and gamma-rays produced in the outermost layers escape, therefore our
simple gamma-ray deposition prescription cannot be used to infer a
nickel mass fraction as a function of velocity, since gamma-ray
deposition is non-local. 

Even though it is likely that there exists a spherically symmetric
deposition function that will better reproduce the observed spectra,
it is clear that the distribution of nickel in SN 1987A is
non-spherical. \citet{chug87a91} found evidence from the He~I
$\lambda10830$ line for a ``jetlike'' ejection of \nni. He found that
he could fit the profile with $10^{-4}$~\msol\ of \nni\ above
5000~\kmps\ and that it was consistent with $10^{-3}-10^{-5}$~\msol\ of
nickel needed to fit the emission spectrum of the hydrogen lines. This
is also consistent with our results. \citet{haas90} observed the [Fe
II] 17.94 and 25.99 $\mu$m lines and found clear asymmetry extending
to about 3500~\kmps, these results were modeled by \citet{nagataki00}
using 2-D hydro and confirm the existence of an asymmetric
distribution of nickel.

\section{Synthetic Light Curves\label{lc}}

\subsection{SEAM Distance\label{mus}}

We have developed an improved version of the ``Expanding Photosphere
Method'' (EPM): the Spectral-fitting Expanding Atmosphere Method
(SEAM).  The EPM method
\citep{baadeepm,branpatchbw73,kkepm} calculates distances to
supernovae by assuming that the bolometric luminosity can be
approximated as
\[ L \approx \zeta^2 4\pi R^2 \sigma T^4, \]
where $\zeta$ is the dilution factor \citep{hlw86a,hlw86b,hw87}, and
represents the fact that flux is diluted since the effective
temperature differs from that of the photospheric temperature (By
energy conservation, $\zeta$ can be both less than one or greater than
one depending on how T is determined). In supernovae, homology is
valid, except during the first instants of the explosion and hence
$R = v t$. The
velocity can be measured from the spectral lines, and $T$ is obtained from
photometric colors. The time since explosion can be determined by
demanding self-consistency among epochs
or from the light curve
\citep{eastkir89,schmkireas92,esk96,hamuyepm01,leonard99em02}. EPM
clearly suffers from the limitation that supernovae are not
blackbodies and spectral line features affect the value of the colors
determined observationally and the theoretical colors that must be
determined from the Planck function to obtain a distance. 
It is also unclear
just which spectral lines should be used to determine the velocity
that is needed to ascertain the radius
\citep{hamuyepm01,leonard99em02}. The other obvious limitation is that
the method depends sensitively on the value of the dilution factors
which are assumed to be a monotonic function of the color temperature.

The SEAM method \citep{b93j1,b93j2,b93j3,b93j4,b94i1} retains the
assumption of spherical symmetry, and homology $R = v t$, but uses
detailed spectral models to calculate a synthetic spectrum.
Then (modulo extinction, see below) since we know the
entire spectral energy distribution, synthetic photometry can be used
to calculate a distance modulus in every photometric band. No
blackbody assumption is required, no dilution factors are needed, and
the velocity is determined by the fit to all the spectral lines, not
just some pre-selected set whose identities may not be certain. A time
series constrains the time of explosion. 

Here, we obtain the radius using the results of the hydrodynamical
model and each of our models predicts the total emitted spectral energy
distribution (SED) at a particular time. We can convolve our
theoretical SED with any filter function and calculate the absolute
magnitude in that band. If we denote the absolute magnitude in band
$X$ as $M_X$ then the distance modulus in a particular band is
$\mu_X$. For every band at every time we can calculate a value of
$\mu_X$ by the relation
\begin{eqnarray}
 \mu_X &=& m_X - A_X - M_X \nonumber\\
  &=& m_X  - (M_X + A_X), \nonumber
\end{eqnarray}
where $m_X$ is the apparent magnitude in band $X$, and $A_X$ is the
extinction in band $X$. We obtain $A_X$ by applying the reddening
law of \citet{card89}  
with $E(B-V)=0.16$ \citep{lunfran96} and $R=3.1$ to the theoretical SED. This
procedure defines the SEAM method. No assumption about the shape of
the SED is made since it is determined by the fit to the observed
spectrum. 

The distance to the LMC is of great importance in establishing the
cosmological distance ladder. The value of the Hubble constant depends
crucially on the calibration of Cepheids \citep{tonry01}, which is
dependent on the 
distance to the LMC. Recent reviews of the distance to the LMC are in
\citet{gibson02} and \citet{feast01}. We focus here specifically on
the distance to SN 1987A, but note that the distance to the center of
the LMC is about $\mu = 0.03 \pm 0.03$ nearer than to the supernova
\citep{mccall93} and that a recent determination to the eclipsing
binary HV 2274 gives $\mu = 18.46 \pm 0.06$ \citep{GS01}.

\begin{figure}
\begin{center}
\leavevmode
\includegraphics[width=14cm,angle=0]{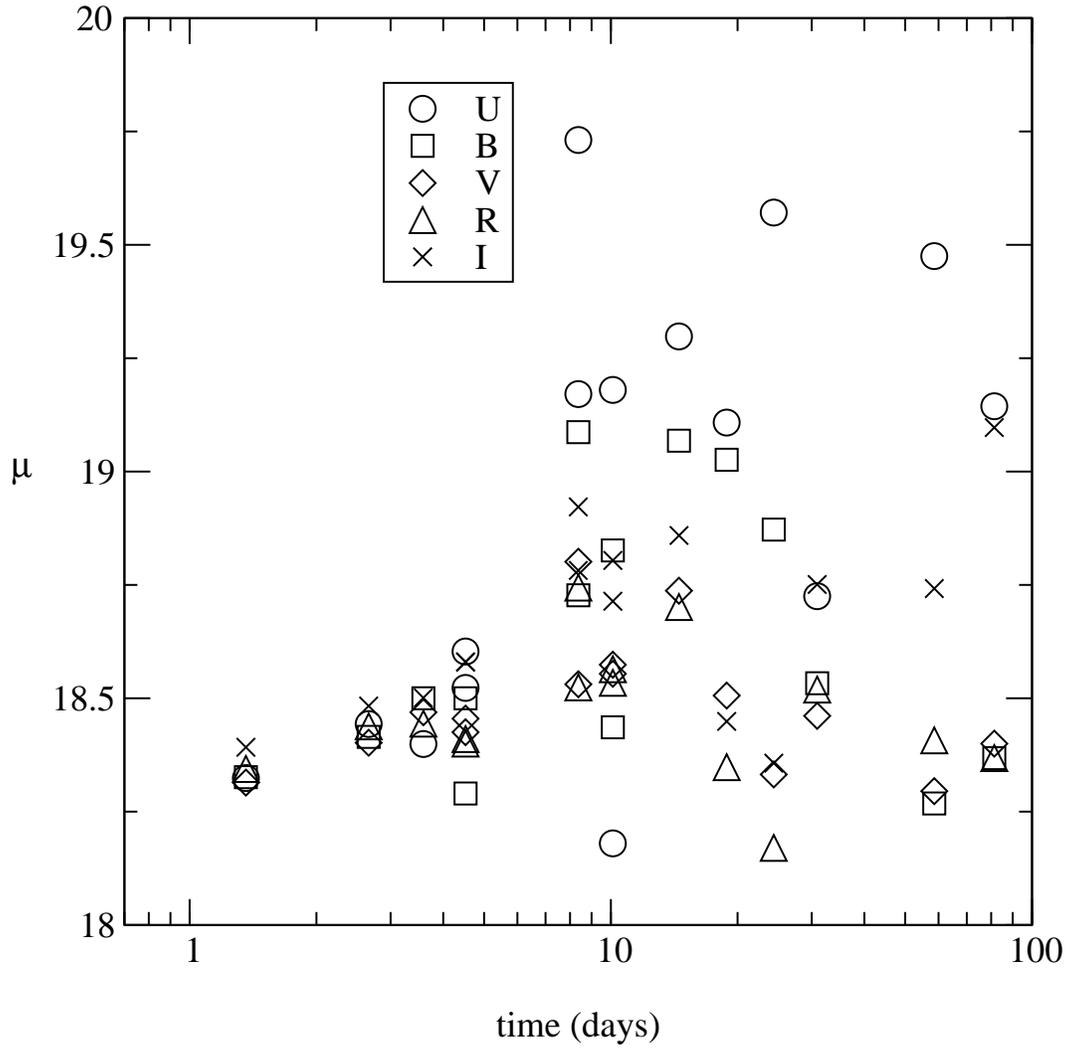}
\end{center}
\caption{The distance modulus obtained from the 15 models we have
calculated. The distance modulus is obtained using the photometry of
\citet{catchpoleetal87A87}, \citet{menziesetal87A87}, and
\citet{HS87A90}. The small spread at early times shows that the SEAM
method is likely to be most accurate at these epochs (see text and
Figure~\protect\ref{fig:seammus}).
\label{fig:mus}}
\end{figure}

Figure~\ref{fig:mus} shows the results for the bands $UBVRI$. The
observed photometry was taken from \citet{catchpoleetal87A87} and
\citet{menziesetal87A87} for the earliest times ($t < 4 $~days), and
from \citet{HS87A90} for all other epochs.  The $U$ band as expected
has a lot of scatter since it is quite sensitive to the assumed nickel
distribution. The mean distance is $\mu = 18.6 \pm 0.3$ using all the
bands. The quoted error here and below is just the 1-$\sigma$ error, 
assuming the errors are Gaussian random errors.  If we limit ourselves to
$VRI$ then we obtain $\mu = 18.5 \pm 0.2$.

Table~\ref{tab:mus} reveals the surprising result that the large
scatter is not limited to $U$ and $B$ but that the $I$ band
shows more scatter than does $R$. Conventional wisdom dictates that
$R$ should not be used for distance determinations because it includes
H$\alpha$, however our results don't bear this out. \citet{hamuyepm01}
found that synthetic $I$ magnitudes on \emph{observed} spectra had
errors at the level of $\sim 0.1$~mag, which may explain, e.g, the
large discrepancy in the value of $\mu_I$ on day 81 in
Figure~\ref{fig:mus}.

\begin{figure}
\begin{center}
\leavevmode
\includegraphics[width=14cm,angle=0]{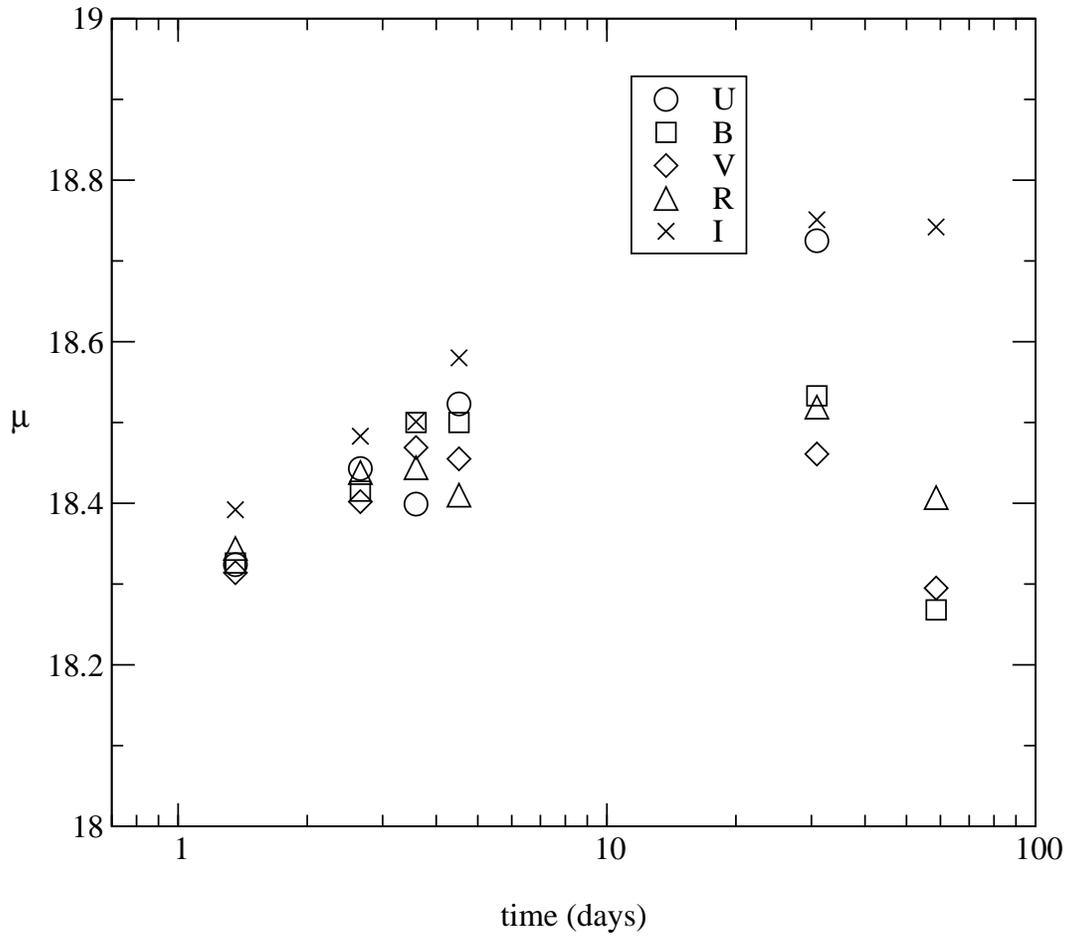}
\end{center}
\caption{The distance modulus obtained from the best fit models we have
calculated. The distance modulus is obtained using the photometry of
\citet{catchpoleetal87A87}, \citet{menziesetal87A87}, and \citet{HS87A90}.
\label{fig:seammus}}
\end{figure}

So far we have used all the model spectra presented in this paper in
determining a distance. The purpose of this is to scrupulously
convince the reader that we have not simply chosen models which
provide us with a desired answer. However, the spirit of the SEAM
method is to use \emph{only} the best fit models, i.e. to use the full
information of the observed SED to determine which models to include
in a distance determination.  If we restrict our distances to just the
earliest times ($t < 5$~days) where the spectra are well fit, we find
$\mu = 18.44 \pm 0.08$ and if we limit our distances to just the best
fit spectra ($t < 5$~days, 31 days, 58 days) we find $\mu = 18.5 \pm
0.2$ using $UBVRI$, and $\mu = 18.46 \pm 0.12$ if we restrict
ourselves to $BVRI$. Figure~\ref{fig:seammus} shows the results for
the SEAM distance, where we require a good fit.  This result shows the
advantage of the SEAM method when the quality of the fit SED to the
data is evaluated to calculate distances. This should tend to reduce
systematic effects when applying SEAM since the sign of the error
varies with both epoch and photometric band. As our final value for
$\mu$ we shall adopt the conservative $\mu = 18.5 \pm 0.2$, but the
least scatter over the widest wavelength range leads to the slightly
lower value of $\mu = 18.46 \pm 0.12$. Table~\ref{tab:mucompare}
compares determinations of the distance to SN~1987A using both EPM and
the ring. Our result is consistent and competitive with even the error
bars of the nearly geometrical ring determination.

While the neutrino signal determines the time of the explosion of
SN~1987A very accurately, it is a useful exercise to examine how the
assumption of the neutrino derived time of explosion affects our
models. From a $\chi^2$ analysis, the neutrino signal time is the
``best'' explosion time in that it minimizes the variance in the
distance modulus, and from the SEAM analysis alone the time is known
to 68\% confidence 
($\Delta\chi^2 = 1$) within $\delta t = \pm 0.9$~days, where the
confidence intervals are symmetric about the time of the neutrino
signal. Thus, the SEAM method is able to determine both the explosion
date and the distance with good accuracy, although our quantitative
uncertainty in the explosion time for SN~1987A likely depends on
having an extremely well sampled early light curve.

\subsection{Light Curves}

\begin{figure}
\begin{center}
\leavevmode
\includegraphics[width=14cm,angle=0]{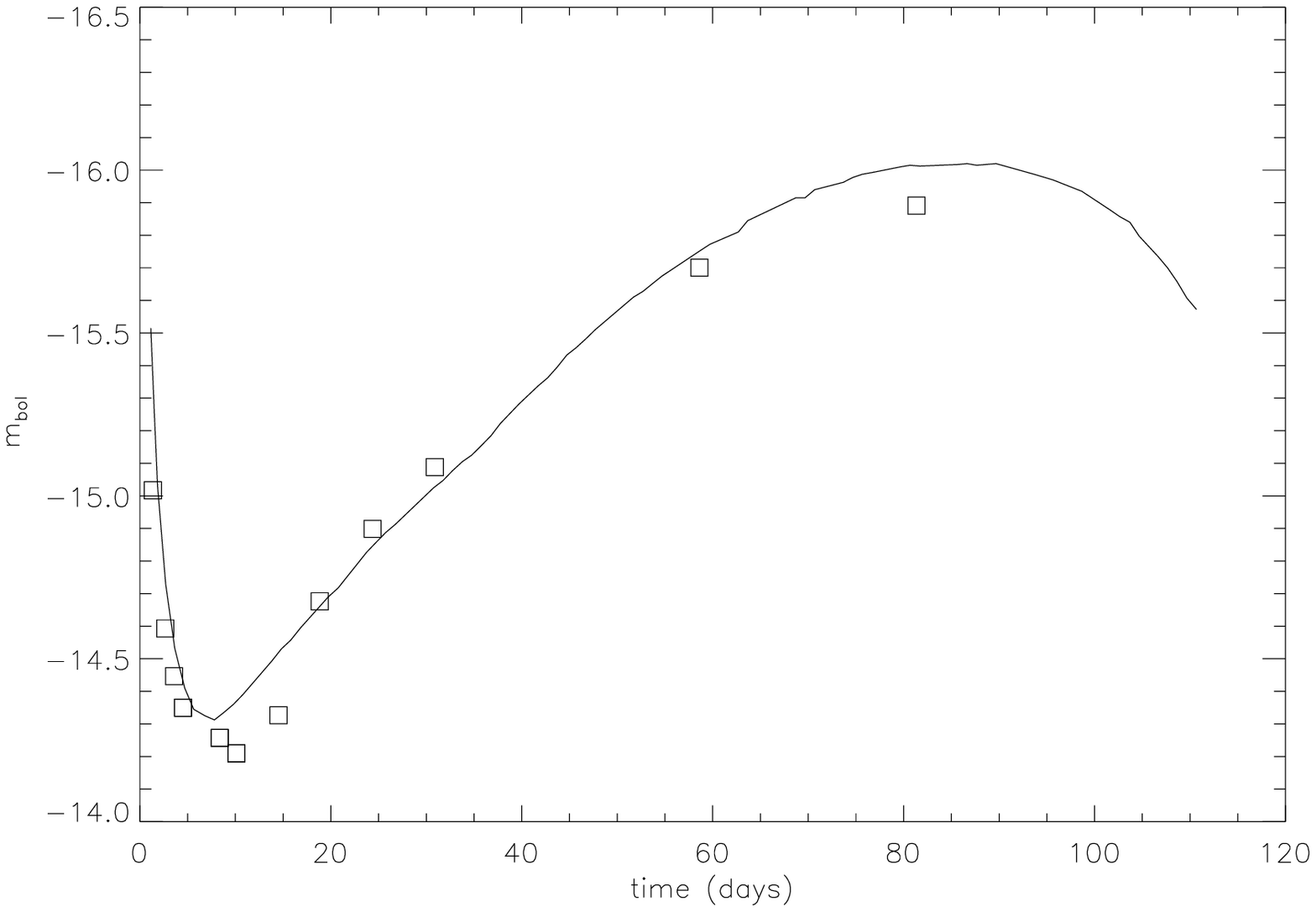}
\end{center}
\caption{The observed bolometric light curve of SN~1987A
\citep[solid line][]{SB87A90} is compared with the model values (open
squares).\label{fig:bollc}} 
\end{figure}

Using our derived value of the distance modulus $\mu = 18.5$, we can
transform our synthetic (reddened) absolute magnitudes into (reddened)
apparent magnitudes. Figure~\ref{fig:bollc} compares our bolometric
light curve to the results of \citet{SB87A90} and
\citet{bouchetetal87A91}. The early times agree quite well,
although the agreement is poor for days 8-24 and then improves
significantly for days $> 24$. This just reflects the fact that the
hydrodynamical model that we have used is not \emph{the} model for SN
1987A. It is clear that days 8-24 should have a higher bolometric
luminosity in order to match the observations, however we have already
increased the bolometric luminosity at these times above that obtained
by \cite{blinn87a00} and larger luminosities than we have used lead to
a significantly poorer fit. Thus, there is not a one-parameter
correspondence between the fit of the observed spectra to the
luminosity.

\begin{figure}
\begin{center}
\leavevmode
\includegraphics[width=14cm,angle=0]{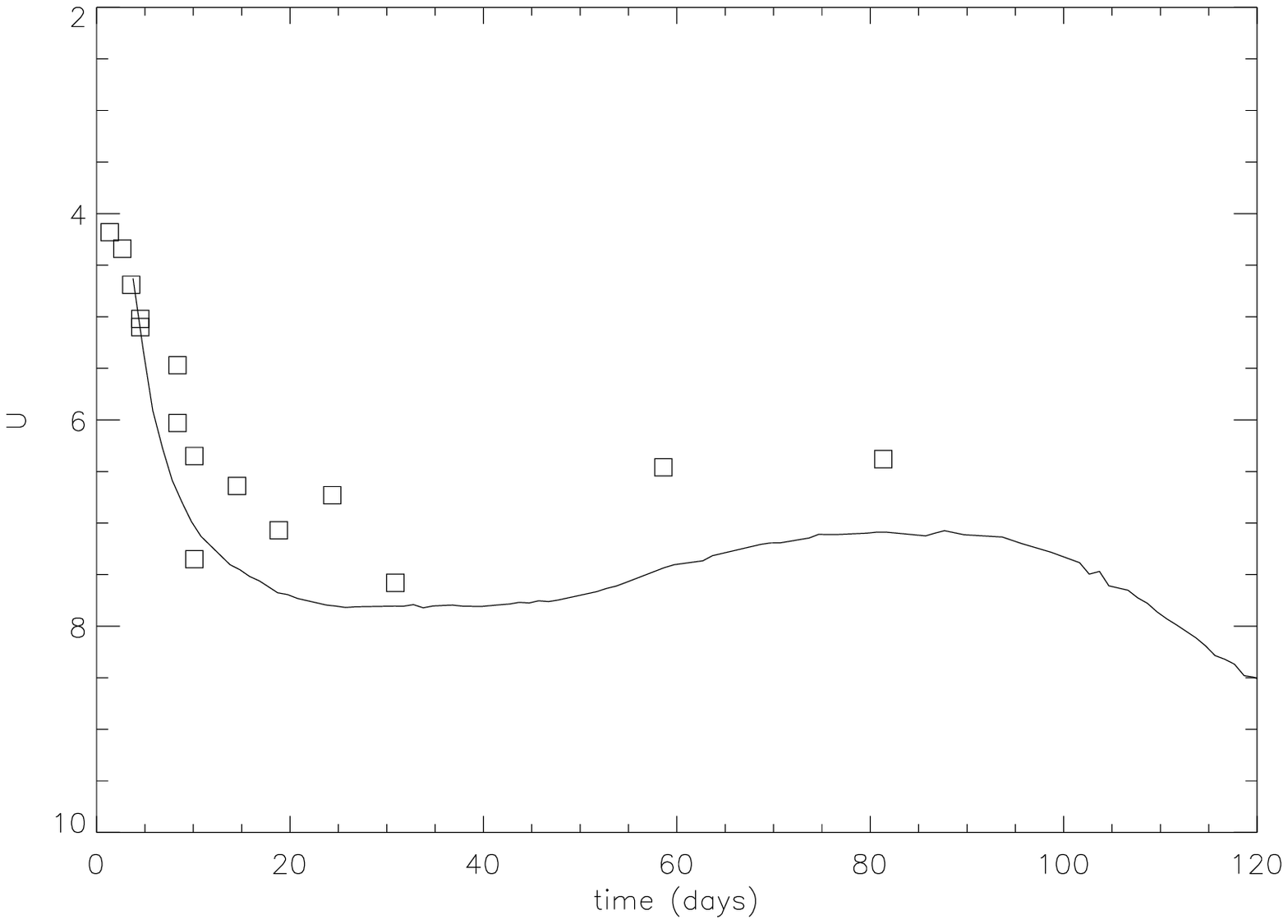}
\end{center}
\caption{The observed $U$ light curve of SN~1987A
\citep[solid line][]{HS87A90} is compared with the model values (open
squares).\label{fig:ulc}} 
\end{figure}

\begin{figure}
\begin{center}
\leavevmode
\includegraphics[width=14cm,angle=0]{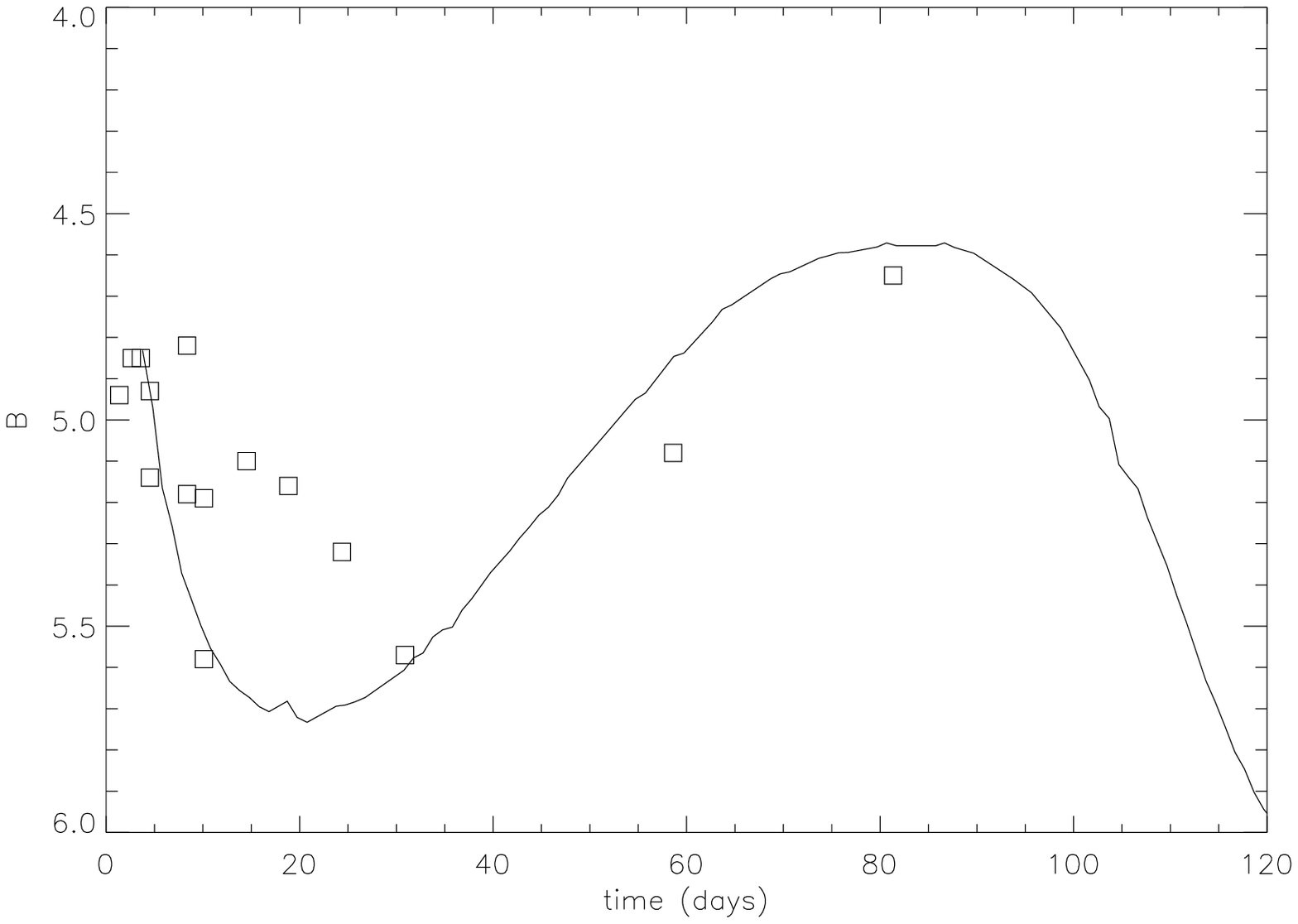}
\end{center}
\caption{The observed $B$ light curve of SN~1987A
\citep[solid line][]{HS87A90} is compared with the model values (open
squares).\label{fig:blc}} 
\end{figure}

\begin{figure}
\begin{center}
\leavevmode
\includegraphics[width=14cm,angle=0]{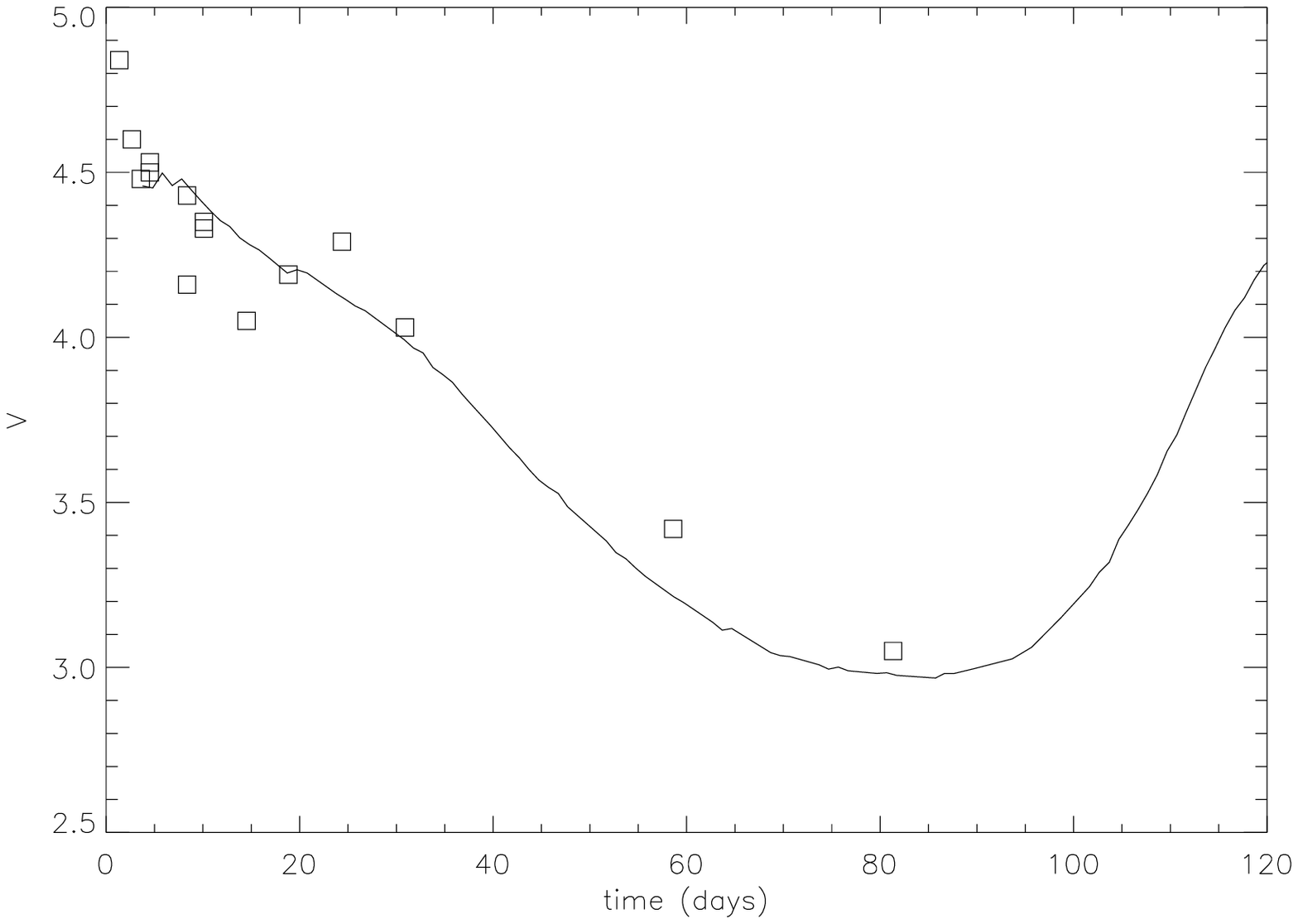}
\end{center}
\caption{The observed $V$ light curve of SN~1987A
\citep[solid line][]{HS87A90} is compared with the model values (open
squares).\label{fig:vlc}} 
\end{figure}

\begin{figure}
\begin{center}
\leavevmode
\includegraphics[width=14cm,angle=0]{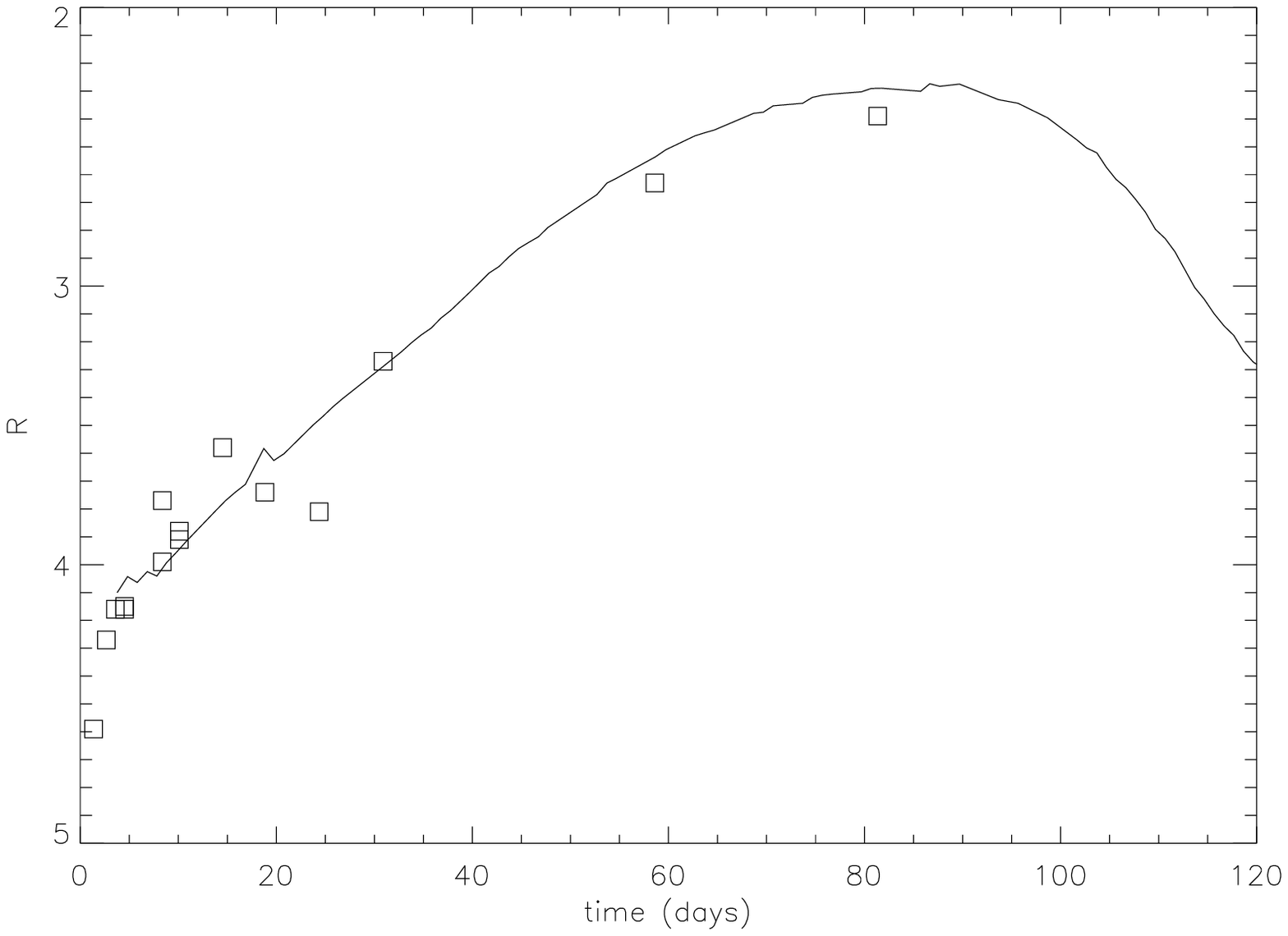}
\end{center}
\caption{The observed $R$ light curve of SN~1987A
\citep[solid line][]{HS87A90} is compared with the model values (open
squares).\label{fig:rlc}} 
\end{figure}

\begin{figure}
\begin{center}
\leavevmode
\includegraphics[width=14cm,angle=0]{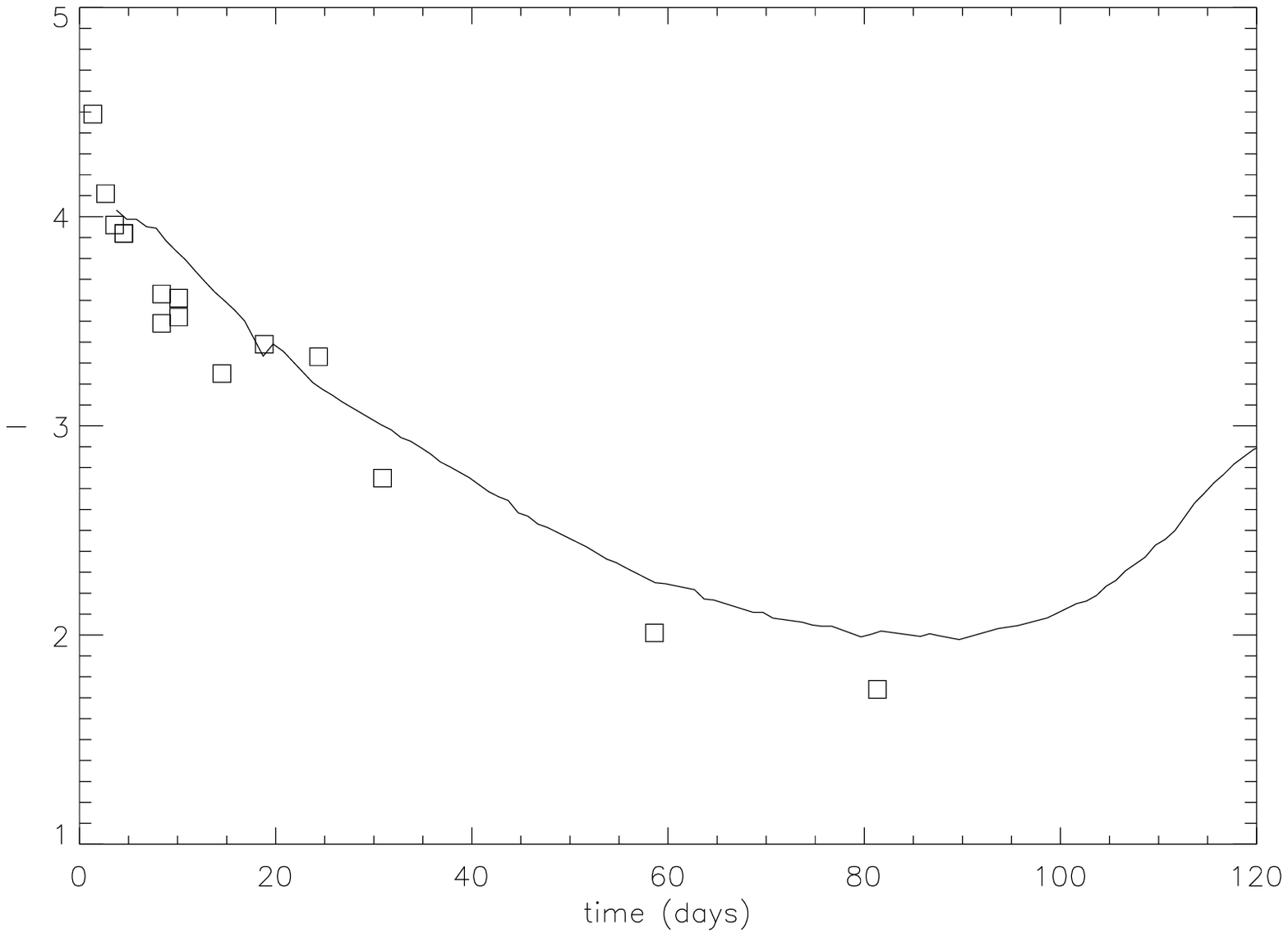}
\end{center}
\caption{The observed $I$ light curve of SN~1987A
\citep[solid line][]{HS87A90} is compared with the model values (open
squares).\label{fig:ilc}} 
\end{figure}

\begin{figure}
\begin{center}
\leavevmode
\includegraphics[width=14cm,angle=0]{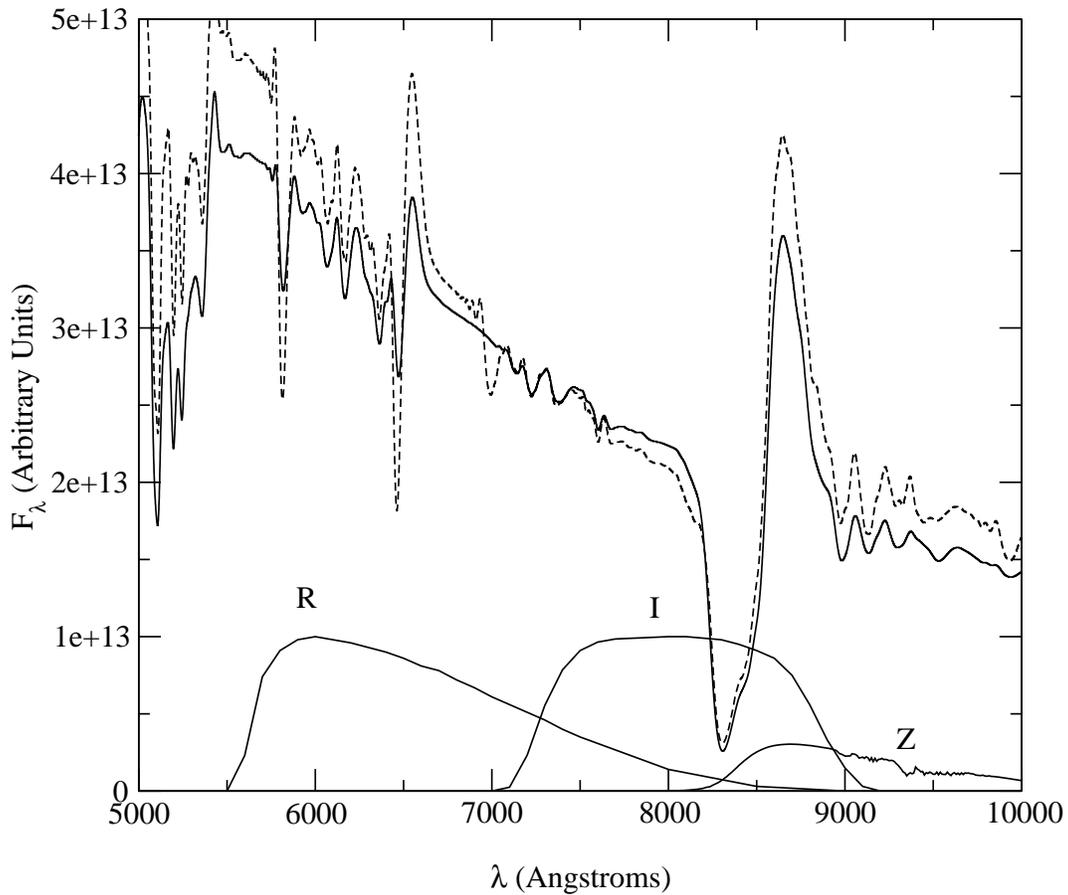}
\end{center}
\caption{Two synthetic spectra for day 18 compared with the filter
response for $R$, $I$, and $Z$. The synthetic spectra differ only in
the total amount of nickel mixing assumed. The solid line corresponds
to the self-consistent deposition function and the dashed line is for
the same shape of the deposition, but multiplied by 100.  Since the
$I$-band doesn't completely cover the Ca IR-triplet, it is very
sensitive to the shape of the continuum around H$\alpha$ and the
IR-triplet. The $Z$-band also is asymmetric about the IR-triplet.
\label{fig:iband}} 
\end{figure}

Figures~\ref{fig:ulc}--\ref{fig:ilc} show the optical $UBVRI$ light
curves compared to the observations \citep{HS87A90}. As we found in
\S\ref{mus}, $V$ and $R$ are most reliable and $UBI$ are less
reliable, in fact $B$ fails completely. The failure in $B$ is
disheartening, but shows that the $B$ band is highly sensitive to the
assumed nickel distribution. Our preliminary calculations show that
early $B$ band is extremely well fit in normal SNe IIP, which we
believe to be better candidates for the SEAM method than are 87A-like
events.  The fact that $I$ is not consistent is largely due to the
lack of observed spectra in the range 7000--9000~\ang. If we had had
spectra to fit this region we would have altered our outer boundary
condition and probably achieved higher accuracy. The fact that $R$
(which includes H$\alpha$) is reasonably well reproduced is
interesting. Figure~\ref{fig:iband} shows that the $I$-band only
covers the absorption component of the Ca IR-triplet and is dropping
in sensitivity in the emission component.  The two synthetic spectra
differ only in the total amount of nickel which clearly affects the
shape of the continuum in addition to that of the lines. The synthetic
$R$ and $I$ both differ by 0.3~mag between the two spectra, but $R-I$
differs only by 0.15~mag, so that EPM analyses that use $V-I$ to determine the
color temperature might be less reliable than using $R-I$. 

As we have already noted the bolometric light curve would be improved
if the models were brighter in the 8-24 day epochs in
particular. However, the $U$ and $B$ band light curves are too bright
in this region. Thus, the density structure of the models must also be
altered in order to match the observed spectra. In particular, since
the blue part of the spectrum forms farthest out (due to
line-blanketing from Fe~II lines) it is possible that the poor fit in
$U$ and $B$ is due to the atmosphere not extending to high enough
velocities, thus there is not enough line blanketing. Line blanketing
leads to both back warming and line cooling and so it may be possible
to reduce the $U$ and $B$ magnitudes, while at the same time
increasing the total bolometric luminosity. It is also possible that
asymmetry is playing a role at these times, but we would expect the
effects of asymmetry to increase with time and so we think that
asymmetry is unlikely to be the cause here. However, the proper
gamma-ray deposition function is certainly needed to reproduce the
blue part of the spectrum.

\begin{figure}
\begin{center}
\leavevmode
\includegraphics[width=14cm,angle=0]{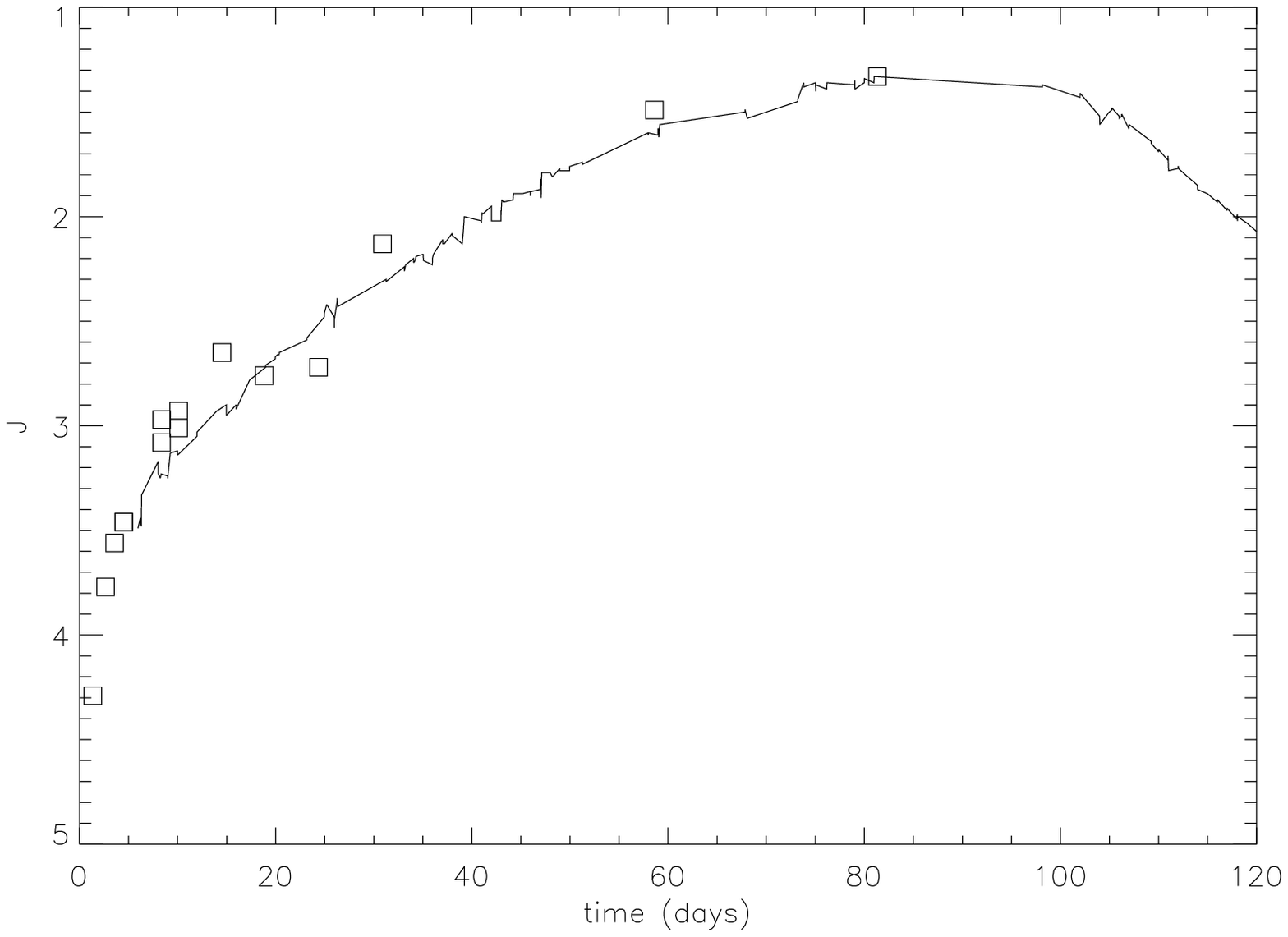}
\end{center}
\caption{The observed $J$ light curve of SN~1987A \citep[solid
line][]{bouchetetal87A89} is compared with the model values (open
squares).\label{fig:jlc}}
\end{figure}

\begin{figure}
\begin{center}
\leavevmode
\includegraphics[width=14cm,angle=0]{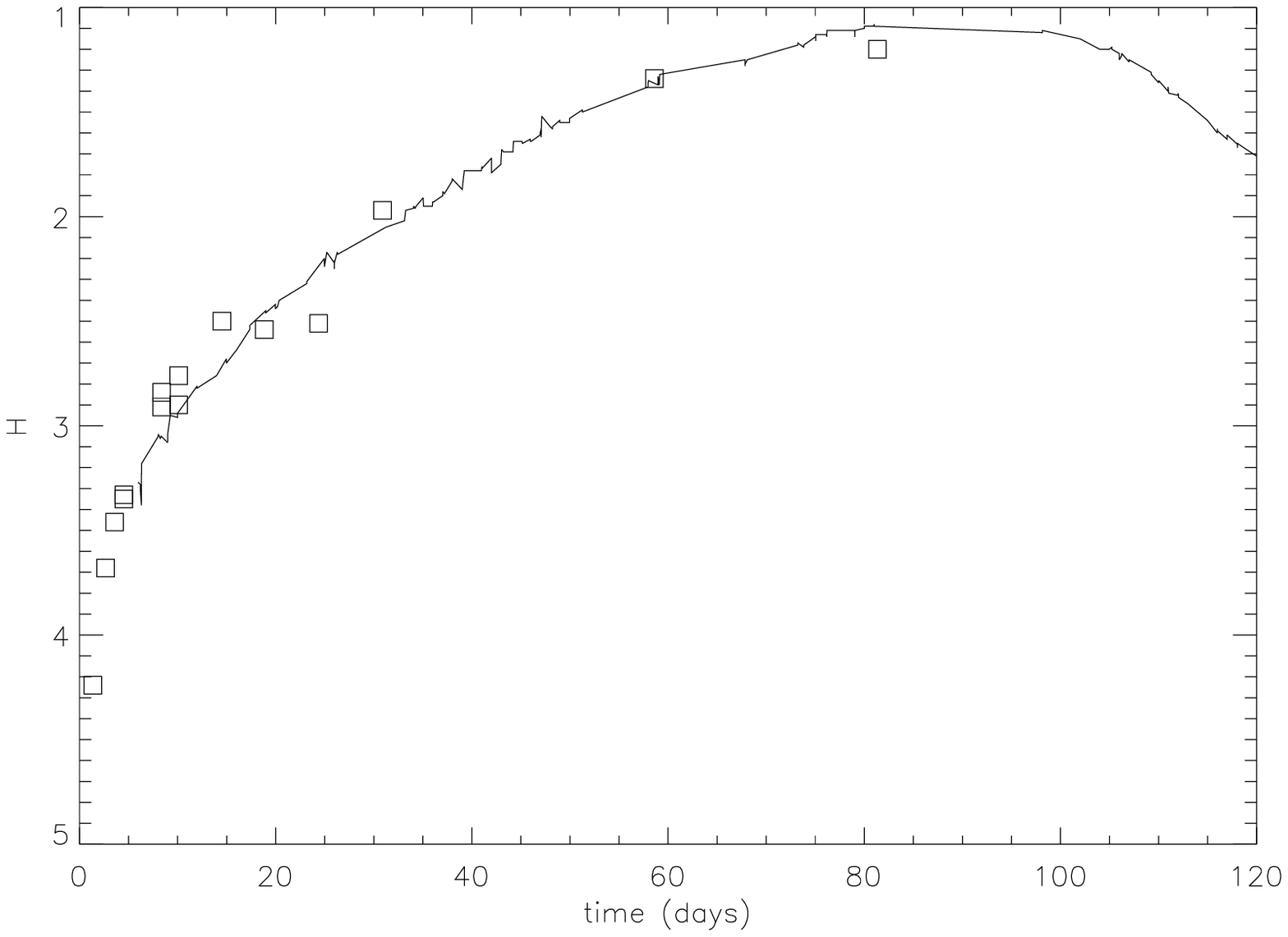}
\end{center}
\caption{The observed $H$ light curve of SN~1987A \citep[solid
line][]{bouchetetal87A89} is compared with the model values (open
squares).\label{fig:hlc}}
\end{figure}

\begin{figure}
\begin{center}
\leavevmode
\includegraphics[width=14cm,angle=0]{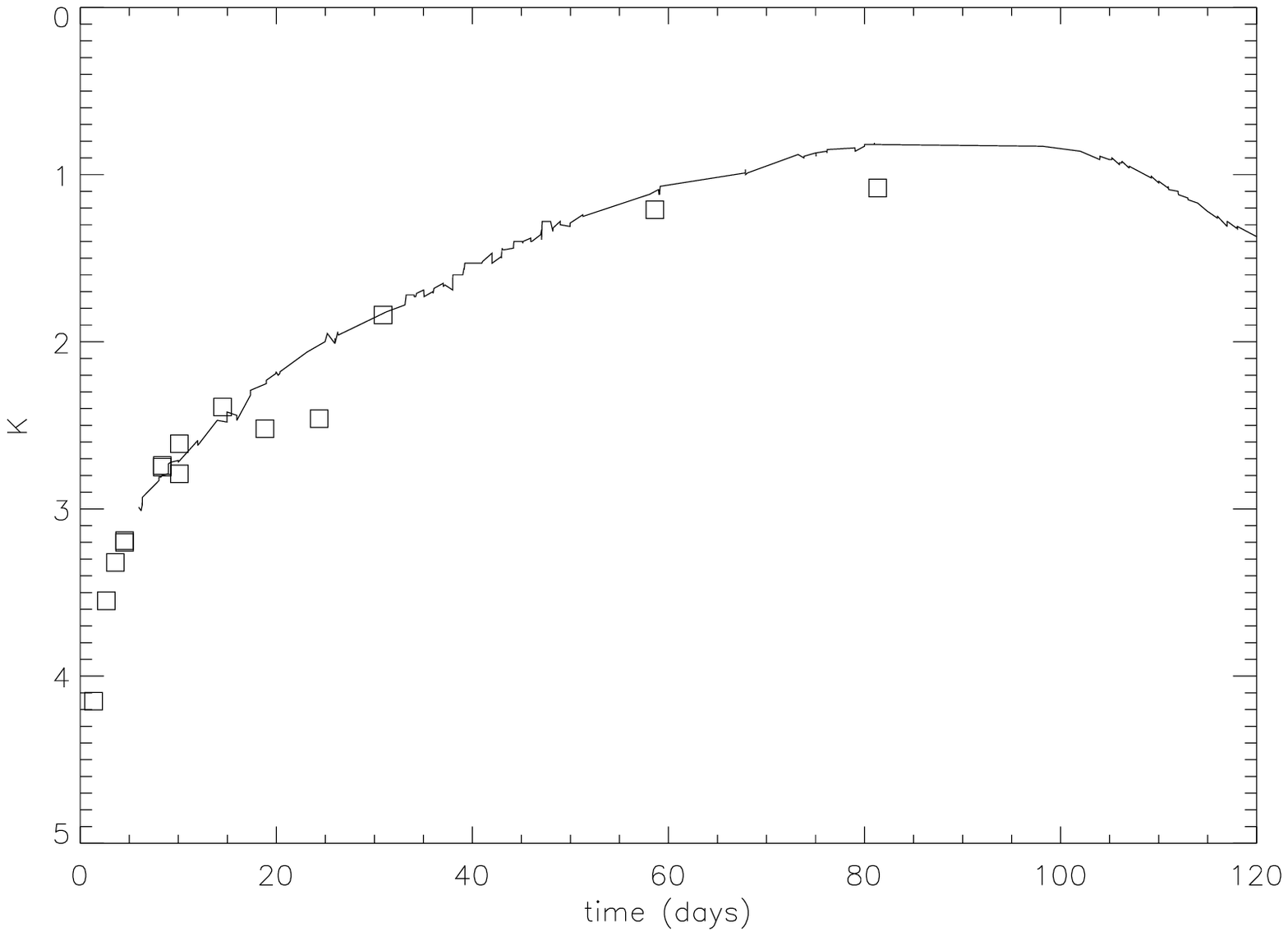}
\end{center}
\caption{The observed $K$ light curve of SN~1987A \citep[solid
line][]{bouchetetal87A89} is compared with the model values (open
squares).\label{fig:klc}}
\end{figure}

\begin{figure}
\begin{center}
\leavevmode
\includegraphics[width=14cm,angle=0]{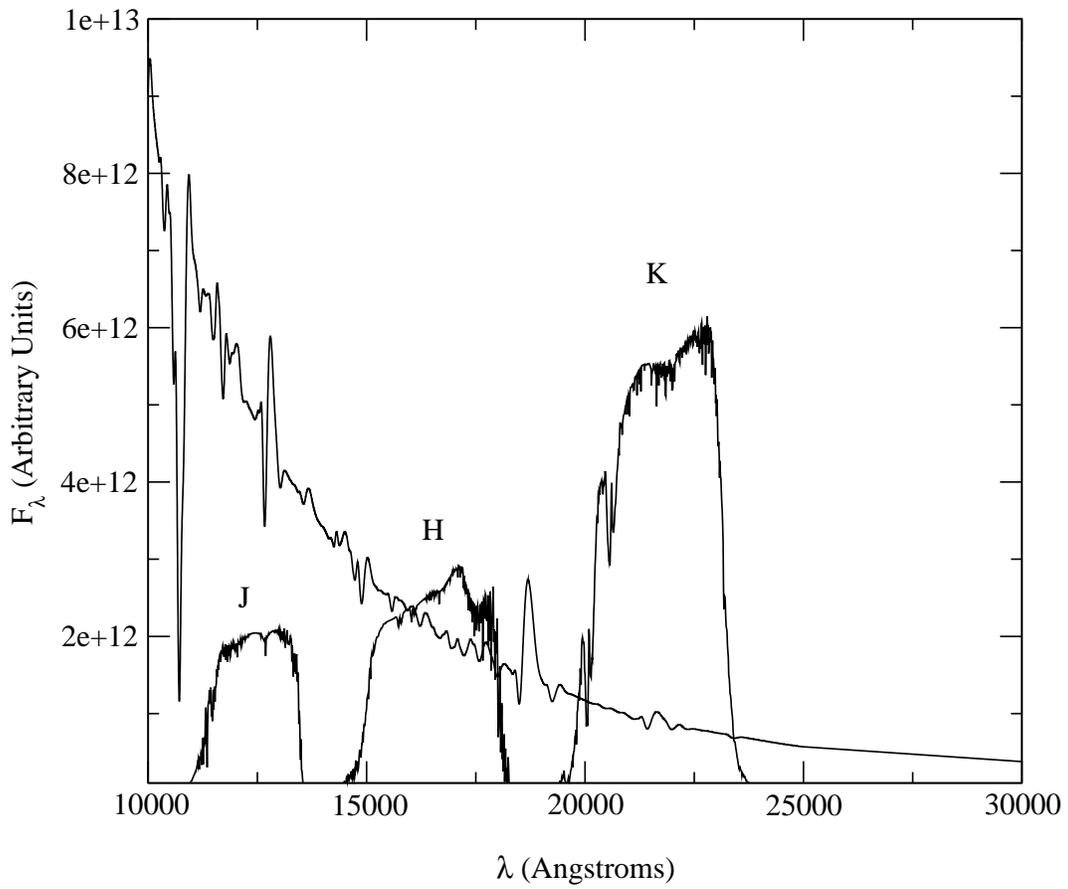}
\end{center}
\caption{The spectrum of the day 30.9 model in the IR, with the
filter+atmospheric transmission for $JHK$. The only prominent lines
are the hydrogen lines P$\alpha$, P$\beta$, and P$\gamma$. Otherwise
the SED is very close to a blackbody.
\label{fig:ir}} 
\end{figure}

Figures~\ref{fig:jlc}--\ref{fig:klc} show the IR ($JHK$) light curves
where even without guidance from observed spectra we obtain very good
agreement. The observations are taken from \citet{bouchetetal87A89}.
This result gives us confidence in the value obtained for 
$\mu$ and it indicates that only small changes are needed in order to
obtain consistent values of $I$. On the other hand it shows that
spectroscopy in the $I$ band 
is required in order to obtain consistent results. Figure~\ref{fig:ir}
shows the IR spectrum of the 30.9 day model with the filter and
atmospheric transmission functions. The SED is nearly a blackbody with
only the hydrogen lines P$\alpha$, P$\beta$, and P$\gamma$ displaying strong
P-Cygni profiles.

\section{Discussion\label{discuss}}

Our results indicate that the SEAM method has promise as an accurate,
competitive distance indicator. It has the advantage over other
methods that its results can be directly compared for consistency and
that the SED includes important effects due to differential expansion
and line blanketing. On the other hand, one would like to see perfect
consistency in every band at every epoch. While 
perfect consistency is a unreachable goal, we expect improved results
with early time normal Type IIP supernovae.
We have used here only a single
hydrodynamical model which is clearly not \emph{the} model of
SN~1987A. We believe that the errors can be significantly
reduced with a large grid of models at each epoch which
we will pursue in future work.   It is particularly
important to have a large range of density profiles for the lower
velocity ejecta in order to correctly reproduce the Balmer lines
consistently over time. It is also important for the model to extend
to the highest velocity seen in the observed spectra.

\begin{figure}
\begin{center}
\leavevmode
\includegraphics[width=14cm,angle=-90]{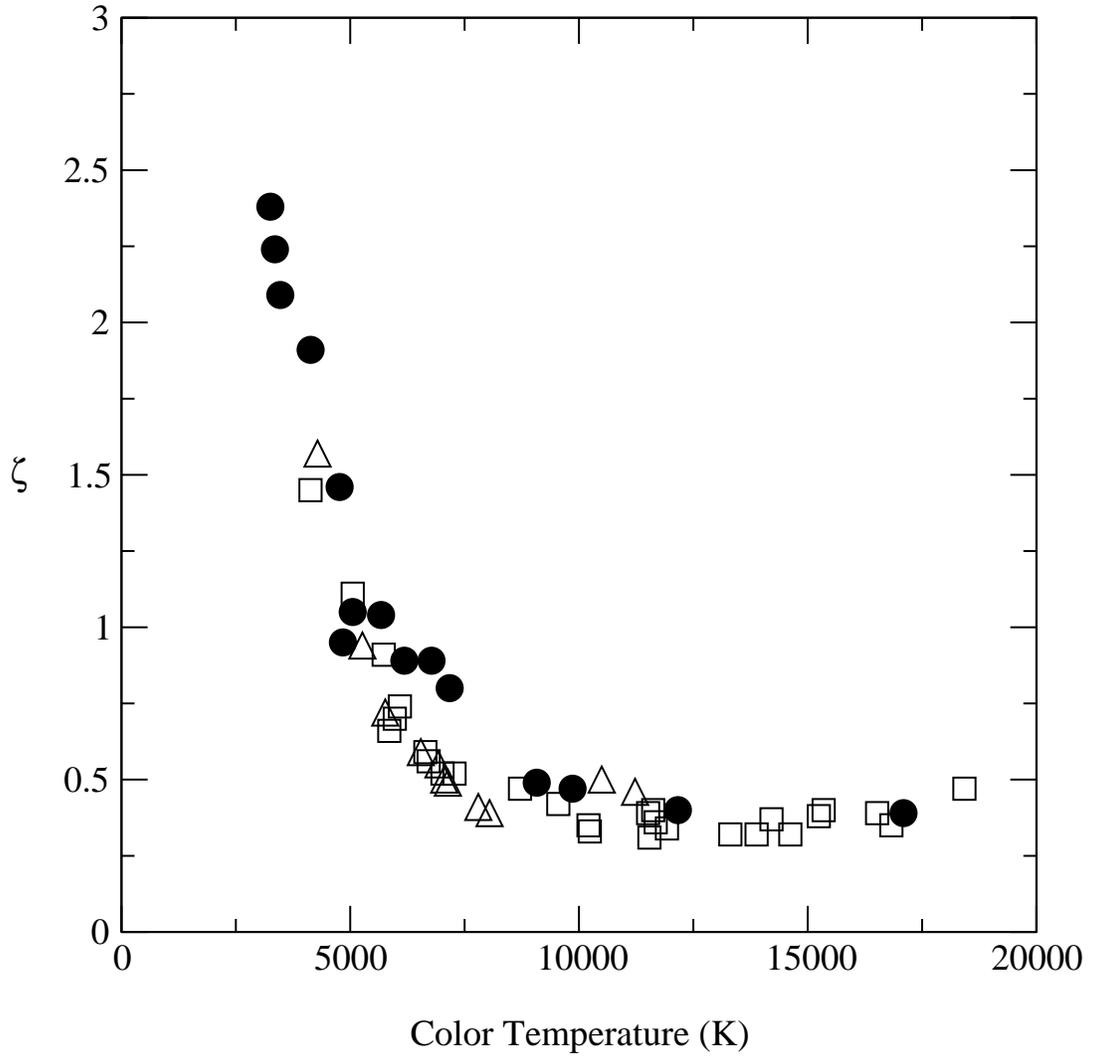}
\end{center}
\caption{The dilution factor $\zeta$ from our models (filled circles)
compared with that obtained by \citet{esk96} for their models s15 (open
squares) and p6 (open triangles).\label{fig:zetas}}
\end{figure}

On the face it may appear that the SEAM method is less consistent than
the EPM method as applied to SN~1999em \citep{hamuyepm01,leonard99em02}.
However, the EPM is more prone to systematic effects than is SEAM,
where the systematic errors are more likely to average out. In EPM the
``temperature'' is obtained from a color temperature from different
filter bands. \citet{hamuyepm01} found that the EPM leads to a
deviation in the distance obtained with epoch (and they did not use
either the $U$-band or $I$-band data). They also found that $BV$ were
less reliable than the redder bands, which is consistent with our
results. \citet{leonard99em02} found that they obtained different
preferred distances when using different filter bands and that
determining the velocities that are required in EPM is
non-trivial. \citet{leonard99em02} found that even obtaining the rest
wavelength of multiplets such as \ion{Sc}{2} is subject to
uncertainty. In reality, there is probably no single unblended line in
the entire spectrum, and the problems of blending become more
important for weak lines that are desired to obtain the photospheric
radius. The need for detailed synthetic spectral modeling is clear. In
fact, the velocity is needed in EPM not only to determine the radius
of the supernova, but also to compare with models in order to obtain
the dilution factor, $\zeta$. Figure~\ref{fig:zetas} shows the results
of our values of $\zeta$ (obtained using $BV$) with that of
\citet{esk96}. The filled circles are our result, the triangles the
power-law model, p6, of \citet{esk96} and the squares the 15~\msol\ 
model, s15, of \citet{esk96}. There does appear to be some difference
between our values and that of \citet{esk96}, particularly for the
important color temperature range of $\sim 5000-8000$~K. Although
\citet{hamuyepm01} found that the power-law model gave more consistent
results for SN~1999em than s15, this cannot be physically correct since real
models must flatten out at low velocity (or they would be
singular).

The fact that the scatter is minimized and the fits are quite good at
early times, leads us to believe that early time normal type IIP
supernovae are likely the best candidates for SEAM. Since polarization
is lowest at early times \citep{leonard01} it is also the time that
the assumption of spherical symmetry is likely to be most valid. Thus we
believe that future SEAM (and EPM) analyses should concentrate on
early time supernovae. It is now up to dedicated supernova searches
to find supernovae at early times and  \emph{to follow them both
photometrically and  spectroscopically}.

\acknowledgments We thank Mario Hamuy for providing us with IR-filter and
transmission functions and for helpful discussions on synthetic
photometry and we thank Claes Fransson for helpful
discussions.   This work was supported in part by NSF grants
AST-9731450, NASA grant NAG5-3505, and an IBM SUR grant to the
University of Oklahoma; and by NSF grant AST-9720704, NASA ATP grant
NAG 5-8425 and LTSA grant NAG 5-3619 to the University of Georgia. SB
is supported in Russia by the grant RFBR 02-02-16500. PHH was
supported in part by the P\^ole Scientifique de Mod\'elisation
Num\'erique at ENS-Lyon. PL acknowledges support from the Royal
Swedish Academy and the Swedish Research Council. Some of the
calculations presented here were performed at the San Diego
Supercomputer Center (SDSC), supported by the NSF, and at the National
Energy Research Supercomputer Center (NERSC), supported by the
U.S. DOE.  We thank both these institutions for a generous allocation
of computer time.


\clearpage

\begin{deluxetable}{cr}
\tablecolumns{2}
\tablewidth{0pc}
\tablecaption{Mean Distance Moduli}
\tablehead{
\colhead{Band}& \colhead{$\mu_X$}}
\startdata
$U$&$18.8 \pm 0.5$\\
$B$&$18.6 \pm 0.3$\\
$V$&$18.5 \pm 0.1$\\
$R$&$18.5 \pm 0.1$\\
$I$&$18.7 \pm 0.2$\\
\enddata
\label{tab:mus}
\end{deluxetable}

\clearpage

\begin{deluxetable}{llll}
\tablecolumns{4}
\tablewidth{0pc}
\tablecaption{Other Distance Determinations}
\tablehead{\colhead{Distance Modulus} &\colhead{Distance (kpc)} &
\colhead{Method}          & \colhead{Source}}
\startdata
$18.7 \pm 0.2$  & $55 \pm 5$     & EPM, $\zeta = 1$  & \citet{bran87a} \\
$18.2 \pm 0.2$  & $43.3 \pm 4$   & EPM             & \citet{chilwag88} \\
$18.45 \pm 0.25$& $49 \pm 6$     & EPM, $\zeta < 1$  & \citet{eastkir89} \\
$18.3 \pm 0.2$  & $45.3 \pm 4$   & EPM, $\zeta < 1$  & \citet{schm90} \\
$18.55 \pm 0.12$& $51.2 \pm 3.1$ & ring (circular) & \citet{nino87aring91} \\
$<18.35 \pm 0.03$  & $< 46.77 \pm 0.76$ & ring (circular) & \citet{gouldring95} \\
$18.53 \pm 0.07$  & $50.9 \pm 1.8$ & ring (circular) & \citet{nino87aring97} \\
$18.43 \pm 0.09$  & $48.6 \pm 2.2$ & ring (circular) & \citet{sonn87a97} \\
$ < 18.53 \pm 0.04$& $< 50.8 \pm 1.0$ & ring (circular) & \citet{goulduza98} \\
$18.44 \pm 0.04$  & $48.8 \pm 1.1$ & ring (elliptical) & \citet{goulduza98} \\
$18.5 \pm 0.24$   & $50 \pm 6$     & ring (circular) & \citet*{gkc87a99} \\
$ < 18.67 \pm 0.08$   & $ < 54.2 \pm 2.2$     & ring (circular) & \citet{ls01} \\
$18.5 \pm 0.2$    & $50 \pm 5$    & SEAM (all data) & This work\\
$18.46 \pm 0.12$  & $49.2 \pm 3$   & SEAM & This work\\
\enddata
\label{tab:mucompare}
\end{deluxetable}

\clearpage

\clearpage

\end{document}